\documentclass[a4paper,english,12pt]{article}
\usepackage[affil-it]{authblk}
\usepackage{babel}
\usepackage{epsfig}
\usepackage{amsmath}
\usepackage{amssymb}
\usepackage[utf8]{inputenc}
\usepackage[T1]{fontenc}
\usepackage[round, authoryear]{natbib}
\usepackage{footnote}
\newcommand{\SortNoop}[1]{}

\newcommand{\E}{\mathbf{E}}
\newcommand{\I}{\mathbf{1}}

\newcommand{\ora}{\overrightarrow}
\parskip\bigskipamount

\usepackage[margin=1in]{geometry}

\usepackage{fancyhdr}
\fancypagestyle{title}{%
  \setlength{\headheight}{22pt}%
  \fancyhf{}
}

\title{Deducing self-interaction in eye movement
data using sequential spatial point processes}
\author[a]{Antti Penttinen}
\author[ab]{Anna-Kaisa Ylitalo\thanks{Electronic address: \texttt{anna-kaisa.ylitalo@jyu.fi}; Corresponding author;}\thanks{Present address: University of Jyvaskyla, Department of Music, P.O. Box 35, FI-40014 University of Jyvaskyla, Finland}}
\affil[a]{Department of Mathematics and Statistics, University of
Jyvaskyla, Finland}
\affil[b]{Department of Music, University of
Jyvaskyla, Finland}

\begin{document}

\maketitle \thispagestyle{title} \pagestyle{title}
\begin{abstract}

Eye movement data are outputs of an analyser tracking the gaze
when a person is inspecting a scene. These kind of
data are of increasing importance in scientific research as well
as in applications, e.g.\ in marketing and man-machine interface
planning. Thus the new areas of application call for advanced analysis tools. Our research objective is to suggest statistical modelling of eye movement
sequences using sequential spatial point processes, which decomposes the variation in data into structural components having interpretation.

We consider three elements of an eye movement sequence: heterogeneity of
the target space, contextuality between subsequent movements, and
time-dependent behaviour describing self-interaction. We propose
two model constructions. One is based on the history-dependent rejection of 
transitions in a random walk and the other makes use of a history-adapted kernel function penalized by user-defined geometric model characteristics. 
Both models are inhomogeneous self-interacting random walks. 
Statistical inference based on the likelihood is 
suggested, some experiments are carried out, and the models are used for determining the
uncertainty of important data summaries for eye movement data. 

\end{abstract}


\clearpage
\section{Introduction}
\pagestyle{plain}

\begin{figure}[!ht]
\centering
  \includegraphics[width=0.5\textwidth]{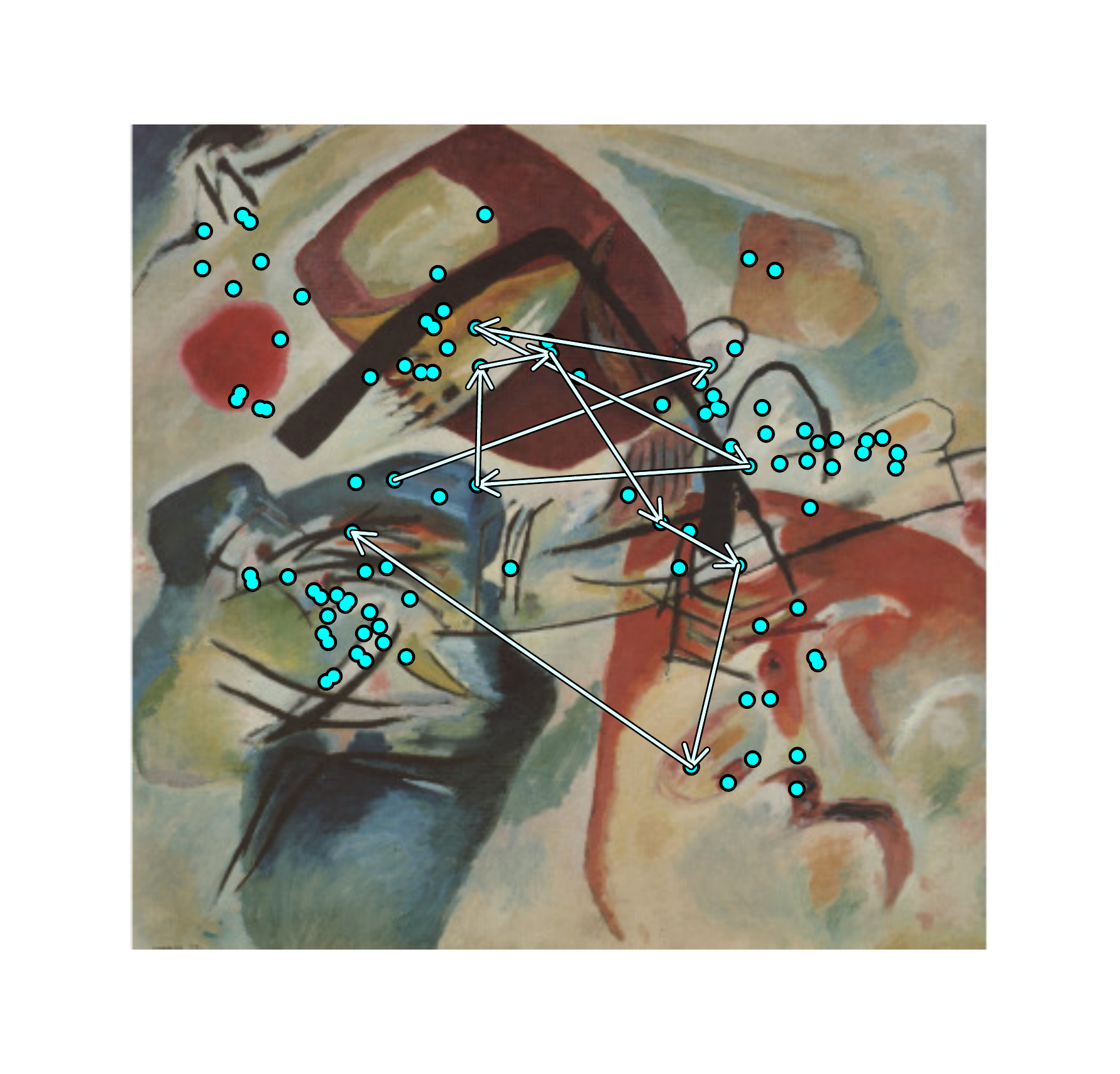}
  \caption{First 100 fixation points of one subject on a painting called Black Bow (1912) by Wassily Kadinsky. The arrows show the movement of the gaze during the first three seconds.}
  \label{fig:Points_kh5}
\end{figure}

Eye movements reflect brain functions, revealing information on ongoing 
cognitive processes, and can be recorded by eye trackers in a cost-efficient 
way. Eye movement data are spatio-temporal and consist of time sequences of
{\em fixations}, points in the target space where the gaze stays for a while,
and of {\em saccades}, which are rapid jumps between fixations. An example of eye movement data can be seen in Figure \ref{fig:Points_kh5}. 

Fixation locations in eye movement data are point patterns which
can be modelled by means of spatial point processes. Point
process statistics is a well-developed branch of spatial statistics increasingly used in
applied sciences, see e.g.\ \citet*{illian_etal} and
\citet*{diggle_2013}. Extensive software \verb+spatstat+
\citep*{baddeley_rubak_turner} has made efficient point pattern data
analysis attractive. Point process statistics has
been applied for eye movement data by \citet*{barthelme_etal}, who use 
the spatial inhomogeneous Poisson point process to predict the 
fixation locations. 
The approach by \citet{barthelme_etal} aggregates the eye movement
data over time but omitting all dynamics. \citet*{engbert_2015} present a dynamical model that takes spatial interaction into account, but their model validation is based on characteristics of spatial point processes. A step towards
a dynamic model is to add the temporal order of
fixations, which leads us to the class of (finite) sequential spatial
point processes, see \citet*{lieshout2006a, lieshout2006b, lieshout2009}.
If, in addition to the order, the time instances of occurrences
of the points are recorded and included in the model, then the underlying process is a spatio-temporal point process, see e.g.\
\citet*{diggle_etal_2010}, and an application to eye movement data by
\citet*{ylitalo_etal}. 

We consider eye movement data to be a realisation of a sequential spatial
point process which allows us to extend the approach by
\citet{barthelme_etal} for detecting new
important dynamic structures of data. The advantage of this approach is that the likelihood is 
tractable and the simulation of realisations is straightforward. 
In addition, sequential point process modelling is a construction
step for spatio-temporal point processes.

For eye movement data, three structural components of sequential spatial point
processes are central. Spatial {\em heterogeneity} of fixation
pattern means that some parts of the scene get the observer's 
attention more than others. This strong component is present in almost all 
eye movement data. It is usually modelled through a saliency map, which is calculated from the features of the scene \citep*[see e.g.][]{itti_koch_2000, kummerer}, such that the most salient areas are expected to obtain more fixations. {\em Dynamic contextuality} is a saccadic property which describes the (metric) length of a jump from the current fixation to the
next one: for example, nearby sites may be more favourable than the more distant ones \citep*[see e.g.][]{tatler_etal_2006}. 
Both heterogeneity and contextuality are well-established in eye movement studies \citep[see e.g.][]{barthelme_etal, engbert_2015, kummerer}. 

However, our empirical evidence shows that these two components are not 
sufficient, since e.g.\ they cannot model the learning effect. Similar findings are made by \citet{engbert_2015} and \citet{kummerer}.
Thus we are looking for simple mechanisms which could utilize the
long-term dependence indicating the learning process during an
experiment. One such possibility is {\em self-interaction}, 
which
modifies the individual moves (saccades) by means of the history
of the eye movement sequence. As an illustration, the observer
prefers to inspect the whole scene at the beginning of the
experiment and gradually focuses on a few details \citep*[see e.g.][]{locher_gray_nodine}. Here, we offer a tool for studying the self-interaction effect in eye movement sequences. 
We suggest new models for eye movement data to deduce the effect of structural components and to evaluate
statistical variation in problem-specific functional summary statistics. The suggested
models have potential use also beyond the eye movement research, e.g.\ in ecology for modelling
animal movements, and in user-interface studies.

Our idea is that the history of the sequence changes the dynamics
during the evolution of the eye movement sequence. We present two
general principles for model construction, both of which
are generalisations of the random walk in heterogeneous media. The first
principle is the history-dependent thinning of transitions, which assigns smaller weights for suggested transitions being at odds
with the chosen functional summary characteristic conditional on
all previous fixations. This type of penalization is similar to the area-interaction process by \citet*{baddeley_lieshout} in point process statistics. The second principle resembles the ARCH
(autoregressive conditional heterogeneity) model, commonly
applied in econometric time series analysis \citep*{engle}.
Based on these principles, we present the construction of the two processes, how to simulate them, and how to estimate the
parameters by the maximum likelihood method. Several
summary statistics, assisted by Monte Carlo simulation, are applied 
in model evaluation. 

The new models are mainly of ``statistical'' nature, which means
that they do not mimic the neural process, but they can capture
essential variation in eye movement data. We will not employ
all the generality the suggested new models are able to achieve.
Instead, the objective is to present the new ideas in terms of
rather simple models which still are useful in eye movement
data analysis, especially in the study of the learning mechanism
during an experiment, and in the derivation of statistical
variation of important data summaries. Furthermore, this new approach will bring eye movement data
analysis closer to statistical inference. The
motivation is the complexity of eye movement data, which are
inhomogeneous in space and time, and the use of asymptotic
inference, for example, is difficult to justify.

The paper is organized in the following way. In Section 2, two new
sequential spatial point process models are suggested. Simulation algorithms are
given in Section 3 with simulation experiments demonstrating the models. 
In Section 4, eye
movement data in the field of art study is modelled by using
the new approach to deduce self-interaction. Section 5 contains some concluding remarks. 

\section{Finite sequential spatial point process models}

Suppose $\ora{x}_n=(x_1,\dots,x_n)$ is a sequence of time-ordered
points in a bounded window $W\subset\mathbb{R}^2$. The corresponding
unordered point set $\{x_1,\dots,x_n\}$ is denoted by
$\{\ora{x}_n\}$. If $(W^n,\mathcal{W}^n)$ stands for the
$n$-dimensional space of ordered points provided with the Borel
$\sigma$-algebra in $W^n$, the density function $f$ w.r.t.\
the Lebesgue measure is defined sequentially as follows:
$f_1(x_1)$ stands for the probability density of the first point,
and the conditional density of a further point $x$ given
$\overrightarrow{x}_k=(x_1,\dots,x_k)$, $k=1,\dots,(n-1)$, is
denoted by $f_{k+1}(x|\ora x_k)$, a
transition probability density for the transition $\ora x_k\rightarrow (\ora
x_k,x)$, and the joint density
of $\ora x_n$ is
\begin{eqnarray}
f(\ora x_n)=f_1(x_1)\,\prod_{k=1}^{n-1} f_{k+1}(x_{k+1}|\ora
x_k)\,. \label{JOINTDENSITY}
\end{eqnarray}
The density $f_1(x)$ can be
assumed to follow a function (e.g.\ the saliency map) exposing the focal areas of the
target, or modelling can
be conditional on the first observation $x_1$. The transition
densities $f_{k+1}(x|\ora x_k),~~ k=1,\dots, n-1,$ reflect
saccadic features of the eye movement sequence.

A simple model for the transition would be a random walk in
heterogeneous media, which is defined as
\begin{eqnarray}
\label{model0} f_{k+1}(x|\ora x_k)\propto \alpha(x)\,K(x_k,x),
\end{eqnarray}
where $\alpha(x)$ is non-negative and bounded in $W$, and
$K(x_k,x)$ is a Markovian kernel, i.\,e., 
\begin{eqnarray*}
K(x_k,x)&\geq& 0 \mbox{ for all $x_k,x\in W$, and}\\
\int\displaylimits_W K(x_k,u)\,{\rm d}u&=&1 \mbox{ for all $x_k\in W$}\,,
\end{eqnarray*}
$k=1,\dots,n-1$. In this simple model, $\alpha(x)$ describes
heterogeneity of the scene. It can be a known saliency map, 
an empirical saliency map estimated as the intensity of repeated fixation patterns \citep*[see the discussion in][]{diggle2007second}, or a model based prediction of the saliency map, for
example,
\[
\alpha(x)=h\left(\sum_{j=1}^p b_j z_j(x)\right),
\]
where the variables $z_j(x)$ are the values of $p$ feature vectors
at $x$ extracted from the scene using machine learning techniques, $b_j$:s are regression coefficients,
and $h$ is an adequate non-negative function
\citep[see e.g.][]{barthelme_etal}. The Markovian kernel $K(x_k,x)$ describes the contextuality of
subsequent fixations related to jump lengths. An example is the
truncated Gaussian kernel 
\begin{eqnarray}
\label{truncnormal} K(x_k,x)\, \propto e^{-\frac
1{2\sigma^2}\,||x_k-x||^2}  \,,~~ x_k,x\in W,\,
\end{eqnarray}
where $||x_k-x||$ is the Euclidean distance between the points $x_k$ and $x$. For the rectangular window $W = [a,b] \times [c,d]$ the normalization term for (\ref{truncnormal}) can be written as
\[
2 \pi \sigma^2 (\Phi(b)-\Phi(a))(\Phi(d)-\Phi(c)),
\]
where $\Phi$ is the c.d.f. of the standard normal distribution. This kernel penalizes large jumps and keeps the process inside the specified window $W$. Thus the model (\ref{model0}) captures heterogeneity in the target and
models the transitions (or saccades) in a Markovian way. 

The transition mechanism (\ref{model0}) may be insufficient if data contain learning. For instance,
a two-stage model describing the nature of an aesthetic experience \citep[see e.g.][]{locher, locher_gray_nodine, locher_etal2007} suggests that a picture is first inspected
globally and, after having obtained a gist of the scene, the viewer starts
to concentrate on some details. Also, the
visual information gathered from the scene affects our cognitive
processes and attention, which again affect the movement of the gaze. By
keeping these complexities of human attention in mind, we develop two models which try to catch
the sequential adaptation in the eye movement sequence in a tractable manner using geometric reasoning.

\subsection{Self-interaction due to history-dependent rejection model}

First, we define a {\em history-dependent rejection model} (later:
rejection model), in which the self-interaction mechanism is
created by a reweighting probability function. This model penalizes the location of the next point in terms of {\em coverage} or {\em recurrence} composed by the previous
points: the density for the transition $\ora x_k\rightarrow (\ora
x_k,x)$ is assumed to be
\begin{eqnarray}
\label{model1} 
f_{k+1}(x|\ora x_k) \propto
\alpha(x)\,K(x_k,x)\,\pi(x,S(\ora x_k,x)),
\end{eqnarray}
where $\pi(x, S(\ora x_k,x))$ is the reweighting probability of $x$ when proposed according to the density proportional to $\alpha(x) K(x_k,x)$. Here, $S(\ora x_k,x) = S(x_1, \dots, x_k, x)$ is a
measure of coverage or recurrence of the ordered sequence
$(x_1,\dots,x_k, x)$. 

Reasonable choices of the reweighting probability $\pi$ in the eye movement context are given below. 

{\bf Coverage-based reweighting}

Two measures of the coverage of a point set are the area of its convex hull and
the area of the associated ball union. From now on
we assume that the scene $W$ is convex. The convex hull of a point
set $\{\ora x_k\}$, denoted by ${\rm Conv}(\ora x_k)$,  is the
minimal convex subset of $W$ which contains all the points of $\{\ora
x_k\}$. The convex hull is unique and invariant under permutation
of the points; hence it is the same for ordered and unordered
sets. Let us denote by $S_C(\ora x_k)$ the
area of ${\rm Conv}(\ora x_k)$ and call it convex hull coverage. 

The ball union measure of a point
set $\{\ora x_k\}$ is defined as
\[
{\rm Bcov}(\ora x_k)=\bigcup\limits_{i=1}^k b(x_i,r)\cap W,
\]
where $b(x,r)$ stands for the ball with radius $r$ and centred at $x$. It is a ``regionalized''
version of the point set, where the $r$ close neighbourhood of a point is taken into account, and which again is
invariant under permutation. Its area $S_B(\ora x_k)$ is called ball union coverage.

The rationale behind model
(\ref{model1}) is that the kernel function generates random jumps
and the reweighting probability determines which of the proposed jumps are
accepted, depending on the current coverage of the sequence and on
the new suggestion. Consider the convex hull coverage first: if
the new suggestion $x$ is not in ${\rm Conv}(\ora x_k)$, the odds
ratio of acceptance w.r.t.\ a proposal $y\in {\rm Conv}(\ora
x_k)$ with $||x-x_k||=||y-x_k||$ is $S_C(\ora x_k,x)/S_C(\ora
x_k)$.

A reasonable and simple choice of the geometric nature for the reweighting probability
would be
\begin{eqnarray}
\label{retire} \pi(x, S(\ora x_k,x))=\left\{
\begin{array}{rl}
1 & \mbox{ if $x\in W\setminus {\rm Conv}(\ora x_k)$}\\
\rho & \mbox{ if $x\in {\rm Conv}(\ora x_k)$}\,
\end{array}\right. ,
\end{eqnarray}
where $\rho\in [0,1]$ and $k \ge 1$. If $\rho = 1$, we have the random walk model. When $\rho < 1$, this choice encourages locations outside the convex hull of
previous points leading to faster coverage. When the convex
hull of the points covers almost the whole scene, the process
behaves like a random walk. The density for the transition $\ora x_k \rightarrow (\ora x_k, x)$ with the
truncated Gaussian kernel is
\begin{eqnarray}
\label{trans1} f_{k+1}(x|\ora x_k) \propto \alpha(x)\, e^{-\frac
1{2\sigma^2}\,||x_k-x||^2}\, (\I_{W\setminus {\rm Conv}(\ora
x_k)}(x)+\rho\,\I_{{\rm Conv}(\ora x_k)}(x)),
\end{eqnarray}
where $k=1,2,\dots,n-1$, and $\I(\cdot)$ is the indicator function. 

The convex hull coverage can be replaced by the ball union
coverage. It is not as sensitive to distant points as the convex hull
coverage but reacts to the ``holes'' in the point pattern. If ${\rm Conv}(\ora x_k)$ is replaced by ${\rm Bcov}(\ora x_k)$ in the reweighting probability (\ref{trans1}), the process
favours locations away from the previous points and hence reduces
clustering if $\rho$ is small. It should be noted that the ball union
coverage measure requires the radius of the
ball.

{\bf Recurrence-based reweighting}

As a measure of recurrence we propose the number of earlier visits
in a ball $b(x,r)$ around a site $x$, formally 
$\tilde{S}_R(\ora x_k,x)=\sum_{i=1}^k\I_{b(x_i,r)}(x)$. However, instead of using all the previous points,  at step $k$ the number of earlier visits is calculated from the point set $\{\ora x_{k-1}\}$ omitting the
most recent point $x_k$. This delayed recurrence measure 
\begin{eqnarray}
S_R(\ora x_{k},x) = \sum_{i=1}^{k-1}\I_{b(x_i,r)}(x)
\label{del_rec}
\end{eqnarray}
is less confounded with the Markovian kernel $K(x_k,x)$ than $\tilde{S}_R(\ora x_k,x)$ and is therefore used in this paper
from now on. Note also that the recurrence measure is not
invariant under random permutation.

A simple model for the reweighting probability is
\begin{eqnarray}
\label{rec_acc} \pi(x, S_{R}(\ora x_{k}, x))=\left\{
\begin{array}{ll}
\theta & \mbox{if $S_R(\ora x_{k},x) \ge 1$}\\
1-\theta & \mbox{if $S_R(\ora x_{k},x) = 0$}\,
\end{array}\right. ,
\end{eqnarray}
where $\theta \in [0,1]$ and $k \ge 2$. The odds ratio for accepting a location
close to the points of $\{\ora x_{k-1}\}$ against accepting a location
from an empty area is $\theta / (1-\theta)$. If $\theta$ is
close to 1, the process favours clustering, and if $\theta$ is small,
the process avoids previously visited local areas around the points $\{\ora x_{k-1}\}$. If $\theta =
0.5$, the process is a random walk. The density for the transition $\ora x_k \rightarrow (\ora x_k, x)$ with the
truncated Gaussian kernel is
\begin{align}
\label{trans12} f_{k+1}(x|\ora x_k) \propto \alpha(x)\, e^{-\frac
1{2\sigma^2}\,||x_k-x||^2}\, ((1-\theta)\I_{\{S_R(\ora x_{k},
x)=0\}}(x)+\theta\,\I_{\{S_R(\ora x_{k}, x)\ge 1\}}(x)), 
\end{align}
where $k=2,3,\dots,n-1$, (and $f_2(x|\ora x_1) = f_1(x)$).

In particular, the  model defined through (\ref{model1}) is among
the simplest ones which satisfy our requirements of self-interacting nature. Note that we need the normalized transition
kernel in the likelihood, because the scaling factor contains
the parameters of the model. The normalizing integral can be
computed using numerical integration. Its
evaluation can be avoided in the simulation of the process,
however.

\subsection{Self-interaction due to history-adapted model}

The motivation behind the {\em history-adapted model} arises
from the saliency map idea and the two-stage model by Locher and colleagues
\citep{locher, locher_gray_nodine, locher_etal2007}. Heterogeneity of the target plays the main
role at an early stage of the process evolution: the areas with high saliency (or intensity) get more fixations than the
low-saliency areas. However, when the target has been inspected well
enough, the jump lengths get
shorter as if the process were mimicking a local inspection process.

This model construction is intended for coverage type self-interaction. We apply directly an adaptive
Markovian kernel $K_{\phi_k}(x_k,x)$, where $\phi_k$ is a function of the points $\{\ora x_k\}$ and determines the width of the kernel. Hence, the kernel changes
in time and affects the jump lengths. The transition probability density can be written as
\begin{eqnarray}
\label{model2}
f_{k+1}(x|\ora x_k)&=& \frac{\alpha(x)\,K_{\phi_k}(x_k,x)}{\int\displaylimits_W \alpha(u)\,K_{\phi_k}(x_k,u){\rm d}u}\\
\label{model2b} \phi_k &=& \phi_k(\ora x_k) \propto H(S(\ora x_k))
\end{eqnarray}
where $H(s)$ is decreasing in $s$. Here $\alpha(x)$ controls
the heterogeneity of the target as in the rejection model, whilst
$H(s)$ models the progress of the coverage. This model resembles the autoregressive conditional
heterogeneity model (ARCH) commonly applied in time series
analysis for modelling volatility \citep{engle}. While ARCH models
use the information from $q$ lagged values, our model is allowed to use the
entire history. Again, this model is a
self-interacting random walk.

In what follows, we make use of the specific model
\begin{eqnarray}
\label{model-spec}
K_{\phi_k}(x_k,x)&\propto&e^{-\frac 1{2\phi_k(\ora x_k)}||x_k-x||^2}\,,\\
\label{model-spec2} \phi_k(\ora x_k)&=& \tau\, e^{-\kappa S(\ora
x_k)/|W|}\,,
\end{eqnarray}
$\tau, \kappa \ge 0$ and $x_k,x\in W$, where $S(\ora x_k)$ is the coverage of $\{\ora x_k\}$. If $\kappa = 0$, the
process is a random walk since the kernel does not change in time. This model contains two parameters, $\tau$ describing the initial kernel width and $\kappa$
modelling the decay as a function of coverage. The transition is determined by the conditional density (\ref{model2}). Both convex hull and ball union coverages are suitable for this construction. 

\subsection{Model fitting and statistical inference}

\subsubsection{Model fitting}
\label{fitting}

We assume that an ordered sequence $\ora x_n=(x_1,\dots,x_n)$ is
observed in $W$. In what
follows we suggest parameter estimation for the two models defined
by (\ref{trans1}) (rejection model), and by (\ref{model-spec}) and (\ref{model-spec2})
(history-adapted model) assuming that the non-negative heterogeneity component $\alpha(x)$ is fixed. In practice, the estimation of $\alpha(x)$ is problematic, as we have pointed out in Discussion.

{\bf History-dependent rejection model}

The log-likelihood for the general rejection model (\ref{model1}) is now a function of the model parameters. For the two parameter model defined by (\ref{trans1}) the
expression
\begin{eqnarray}
\label{LL1} l(\sigma^2,\rho)&=& \sum_{k=1}^{n-1}
\log(\alpha(x_{k+1}))-\frac
1{2\sigma^2}\sum_{k=1}^{n-1}||x_k-x_{k+1}||^2 \\&+&
\log(\rho)\,\sum_{k=1}^{n-1} 1_{{\rm Conv}(\ora x_k)}(x_{k+1}) \nonumber\\
&-&\sum_{k=1}^{n-1}\log\int\displaylimits_W\alpha(u)\,e^{-\frac
1{2\sigma^2}\,||x_k-u||^2}\, (1_{W\setminus {\rm Conv}(\ora
x_k)}(u)+\rho\,1_{{\rm Conv}(\ora x_k)}(u)) \,{\rm d}u\,
\nonumber
\end{eqnarray}
is obtained. Here we use the convex hull coverage in the reweighting probability, but also the ball union coverage could be used. The log-likelihood function for the rejection model
with recurrence (\ref{trans12}) is shown in Section \ref{fitting_rej_model}, formula (\ref{LL3}). The logarithm of the normalizing factor (the last line of (\ref{LL1})) can be computed by numerical integration. The optimization of $l(\sigma^2,\rho)$ w.r.t.\ $\sigma^2$ can be
conducted using numerical optimization, or alternatively, one can
solve the exponential family likelihood equation
\[
\sum_{k=1}^{n-1}\E_{\sigma^2,\rho}\left(||x_k-U||^2 \,|\ora x_k
\right)= \sum_{k=1}^{n-1}||x_k-x_{k+1}||^2\,,
\]
where the expectation is over the conditional distribution of a
new random point $U$ from the distribution $f_{k+1}(x|\ora x_k)$. This can be computed using Monte Carlo maximum likelihood
(MCMCML, see \cite*{geyer}). 

Maximizing the log-likelihood (\ref{LL1}) (or (\ref{LL3})) is costly due to the normalizing integral, and we have chosen to use the profile likelihood approach \citep*[see e.g.][p. 127]{davison}: First, $\sigma^2$ is solved for fixed $\rho$ resulting
in $\widehat{\sigma^2(\rho)}$. Then, $\sigma^2$ in the
log-likelihood is substituted by $\widehat{\sigma^2(\rho)}$ giving the log-profile likelihood
$l_P(\rho)=l(\widehat{\sigma^2(\rho)},\rho)$, which is then
maximized w.r.t. $\rho$ and $\hat\rho$ is obtained. Finally, we fix $\rho = \hat{\rho}$ and compute the corresponding value of $\sigma^2$, namely
$\hat{\sigma}^2=\widehat{\sigma^2(\hat\rho)}$. These steps of coordinate descent are iterated to obtain the maximum likelihood estimates. Alternatively, numerical optimization methods can be used.

{\bf History-adapted model}

The kernel width $\phi_k$ of the $k$th transition of the random
walk is a function of the model parameters and is adapted to the history of the sequence through the coverage measure $S(\ora
x_k)$. The
log-likelihood for the general model given by (\ref{model2}) and (\ref{model2b})
is
\begin{eqnarray}
\label{CH} \sum_{k=1}^{n-1}\left[\log(\alpha(x_{k+1})) + \log
K_{\phi_k}(x_k,x_{k+1})- \log \int\displaylimits_W \alpha(u) K_{\phi_k}(x_k,
u)\,{\rm d}u \right]. \nonumber
\end{eqnarray}
In the special case of (\ref{model-spec}) and (\ref{model-spec2}) the
log-likelihood is
\begin{eqnarray}
\label{LL2} l(\tau,\kappa)&=& \sum_{k=2}^{n-1}\log(\alpha(x_{k+1}))
- \sum_{k=2}^{n-1} \frac 1{2\phi(\tau,\kappa)} ||x_k-x_{k+1}||^2 \\ &-&
\sum_{k=2}^{n-1}\log\,\int\displaylimits_W\alpha(u)e^{-\frac
1{2\phi(\tau,\kappa)}||x_k-u||^2}\,{\rm d}u \nonumber
\end{eqnarray}
with $\phi_k(\kappa,\tau)=\tau\, e^{-\kappa S(\ora x_k)/ |W|}$, $k=1,\dots, n-1$.

The log-likelihood can again be maximized directly, or alternatively,
the likelihood equations can be derived and solved: the estimation
equations are
\begin{eqnarray}
\sum_{k=2}^{n-1}\E_{\tau, \kappa}\left(||x_k-U||^2\,|\ora
x_k\right)/\phi_k&=&
\sum_{k=2}^{n-1}||x_k-x_{k+1}||^2/\phi_k\nonumber\\
\nonumber
\sum_{k=2}^{n-1}\E_{ \tau, \kappa}\left(S(\ora
x_k)\,||x_k-U||^2\,|\ora x_k\right)/\phi_k&=&
\sum_{k=2}^{n-1}S(\ora x_k)\,||x_k-x_{k+1}||^2/\phi_k\,,
\end{eqnarray}
where the expectations are over the conditional distribution $f_{k+1}(x|\ora x_k)$ with parameters $\tau$ and $\kappa$. The estimation
equations are in accordance with the maximum likelihood equations
for the exponential family of distributions.

\subsubsection{Model evaluation}

Model evaluation of spatial dynamic models is typically based on the selected functional
summary statistics which measure different features of the model,
such as coverage, recurrence and jump length as a function of
time (or order). In addition, a saliency map is plotted together with the fixation locations. The random
variation of the summary statistics is estimated from simulations.

Model evaluation is done by estimating several summary statistics from data and plotting the estimates together with the model based simulated pointwise envelopes being a parametric bootstrap method \citep*[see e.g.][p. 53]{efron}. These envelopes indicate statistical variation in the summary statistic under the parametric model assumption. It should, however, be noted that when using the pointwise envelopes as statistical
tests, the multiple testing problem is present and the interpretation
of the envelopes must be done with care, see the discussions in
\cite*{grabarnik} and in \cite*{baddeley_etal_2014}.

Model evaluation is illustrated in the examples in Sections
3 and 4.

\section{Simulation experiments}

Realisations from the suggested models can be simulated
sequentially using conditional distributions
(\ref{model1}) and (\ref{model2})-(\ref{model2b}). We recommend to use the scaled heterogeneity $\alpha(x)/\max_{u\in
W}\alpha(u)$ as the distribution for the first location, or alternatively, to
condition to the first location $x_1$ of data. Assume
that $(x_1,\dots,x_k)$ are simulated. A simple algorithm for
adding a point $x$ to $\ora{x}_k=(x_1,\dots,x_k)$, or
equivalently, simulating from the distribution having density $f_{k+1}(x|\ora
x_k)$, is to apply the accept-reject algorithm, see e.g.\ \citet*[p. 61]{ripley}, which provides an upper bound for the conditional
density. For the two models, the algorithm is as follows:

\begin{itemize}
\item History-dependent rejection model: A point  $x$ following the
conditional density $\alpha(x)\, K(x_k,x)$ is proposed using
the accept-reject method and the proposal is accepted with the reweighting probability $ \pi(x,S(\ora x_k,x))\,. $

\item History-adapted model: The kernel width $\phi_k=H(S(\ora x_k))$ is computed and proposals
from the unnormalized transition density $\alpha(x)\,
K_{\phi_k}(x_k,x)$ are drawn using the accept-reject method.
\end{itemize}

In the following simulation experiment we generate realisations
of the new models with three parameter values in order to
demonstrate the time evolution of the new processes and to see how data summaries capture their properties. We illustrate to what extent and how fast these processes can cover the target area, as well as whether the process starts to cluster in space. Each realisation consists of 100 points located in the
unit square window. In this illustration, for the sake of simplicity, we assume that the
target space is homogeneous, setting $\alpha(x) \equiv 1$, and hence the first
point is drawn uniformly and the same starting point is used for all
realisations. The first sampled point is $x_1 = (0.22,0.41)$.

\subsection{History-dependent rejection model}

First, we demonstrate the history-dependent rejection model with
self-interaction defined through the convex hull coverage,
where the points outside the convex hull of the current point set
are favoured according to the reweighting probability (\ref{retire}). Second, we demonstrate the
rejection model with recurrence self-interaction by using the reweighting probability (\ref{rec_acc}), which takes the number of
generated points near the suggested point into account.

\subsubsection{Coverage self-interaction}
The purpose of this example is to illustrate self-interaction
caused by the parameter $\rho$ in the history-dependent rejection
model with convex hull coverage (\ref{trans1}). We fix the parameter $\sigma^2
= 0.3$ of the truncated Gaussian kernel (\ref{truncnormal}) and
vary the parameter $\rho$:
\begin{itemize}
\item {\em Model a}, $\rho = 1$ (random walk without
self-interaction). \item {\em Model b}, $\rho = 0.1$ (fast
coverage), which accepts points inside the convex hull of previous
points with low probability. \item {\em Model c}, $\rho = 0.5$
(mild coverage), which accepts points inside the convex hull of
previous points with mild probability.
\end{itemize}

\begin{figure}[!ht]
\centering
  \includegraphics[width=1\textwidth]{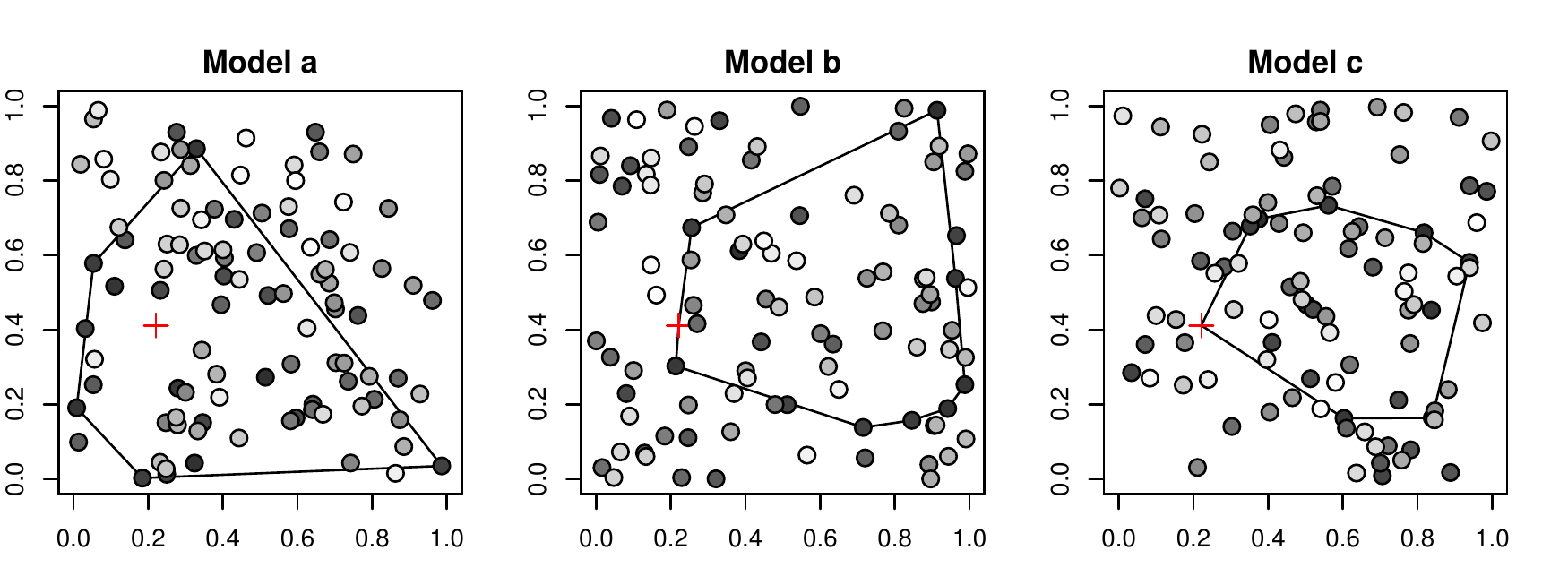}
  \caption{Simulated patterns of the three history-dependent models with coverage self-interaction. The colour of the points denotes their order (from dark to light) and the first fixation is marked with a cross. The polygon indicates the convex hull of the first 10 points.}
  \label{fig:PointsModel1_cov}
\end{figure}

We simulate 19 realisations of the random walk model {\em
a}, since it here represents a reference model, and five
realisations of Model {\em b} and Model {\em c}. One of the
simulated realisations of each model can be seen in Figure
\ref{fig:PointsModel1_cov}. The polygons in the figure
illustrate the convex hull coverage related to the first 10
points of the process. One cannot detect much difference between the
random walk model {\em a} and mild coverage model {\em c}, but the
fast expansion of Model {\em b} can perhaps be seen: there are early (dark) points near the edges.  

\begin{figure}[!ht]
\centering
  \includegraphics[width=0.7\textwidth]{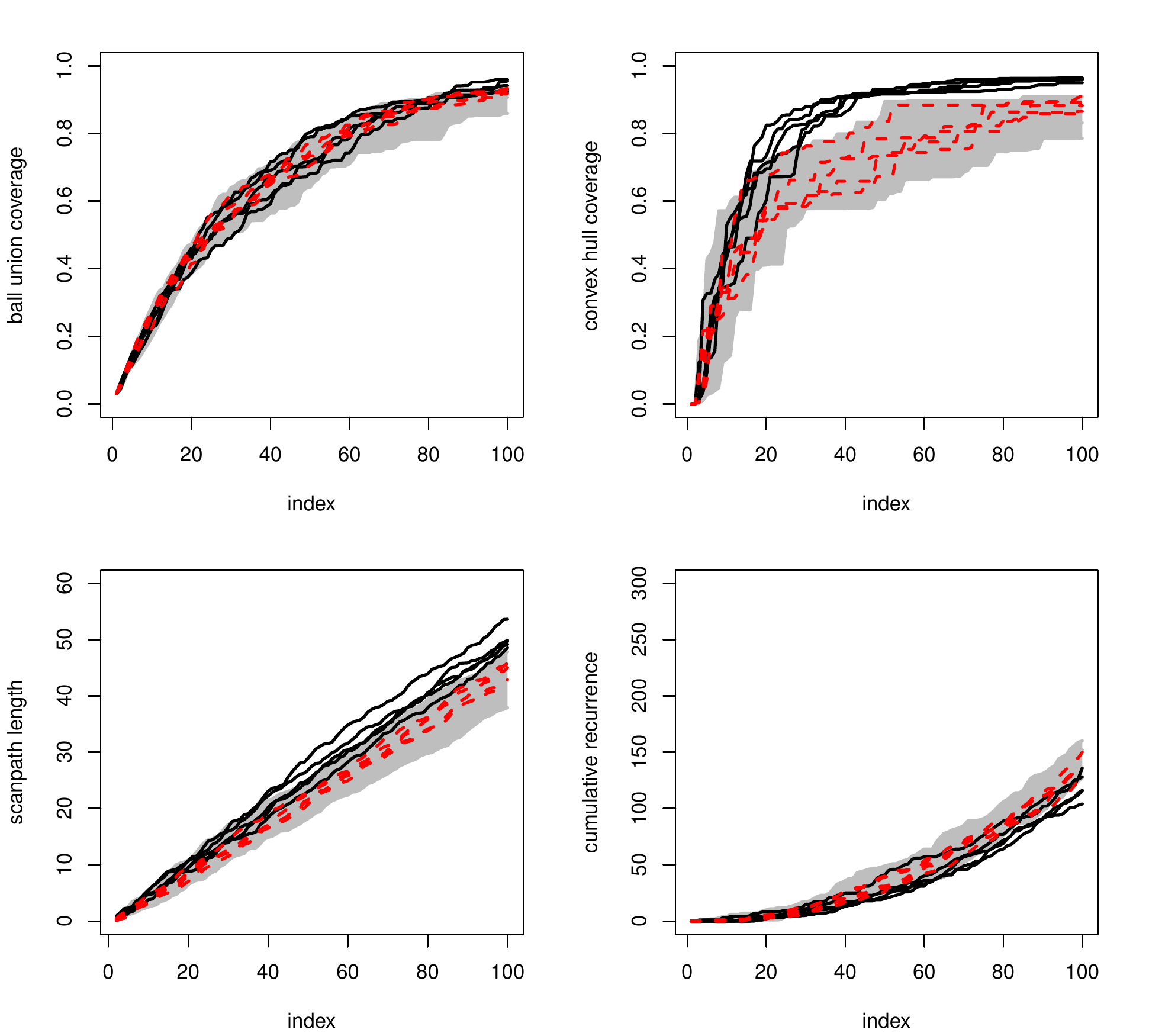}
  \caption{Ball union coverage with radius 0.1 (top left), convex hull coverage (top right), scanpath length (bottom left) and cumulative recurrence with radius 0.1 (bottom right) for the two models (five realisations of each):  The fast coverage model {\em b} is marked with black solid lines and the mild coverage model {\em c} is marked with red dashed lines. The grey area represents the envelopes estimated from 19 realisations of the random walk model {\em a} used as the reference model.}
  \label{fig:SummariesModel1_cov}
\end{figure}

However, point patterns only tell us about the spatial nature of
the point process. Since we are mainly interested in the sequential
(time order) aspect, we use four different functional summary
statistics: ball union coverage (with radius 0.1), convex hull coverage, scanpath
length and cumulative recurrence. (The two latter ones are explained below). The results related to these
summaries can be found in Figure \ref{fig:SummariesModel1_cov}. The
ball union coverage does not distinguish between the three models,
but the convex hull coverage reveals that the coverage of Model
{\em b} increases faster than for the other two models. Accordingly, Model {\em b} makes longer jumps on average
than Model {\em a} or {\em c}, as can be seen from the scanpath length which measures the
length of the sample path cumulatively.

The recurrence function used in the reweighting probability
(\ref{rec_acc}) calculates the numbers of points near the
current point excluding the previous point, and the cumulative
version sums all these numbers together. Now we see that Model {\em b} avoids the
locations nearby other points when compared to the random walk. This may be due to the fact that
the fast coverage process {\em b} penalizes slow coverage and hence increases the drift of points near the edges. 

\subsubsection{Recurrence self-interaction}
Here we illustrate self-interaction
in the history-dependent rejection model with recurrence (\ref{trans12}). We again
fix the parameter $\sigma^2$ of the truncated Gaussian
kernel (\ref{truncnormal}) to 0.3 and vary the self-interaction parameter $\theta$:
\begin{itemize}
\item {\em Model d}, $\theta = 0.5$ (random walk without
self-interaction). \item {\em Model e}, $\theta = 0.1$ (low
recurrence), which favours points in the non-visited areas
rather than the points in the areas nearby the previous points. \item {\em Model f}, $\theta =
0.9$ (high recurrence), which accepts points nearby the previous
points with high probability. 
\end{itemize}

\begin{figure}[!ht]
\centering
  \includegraphics[width=1\textwidth]{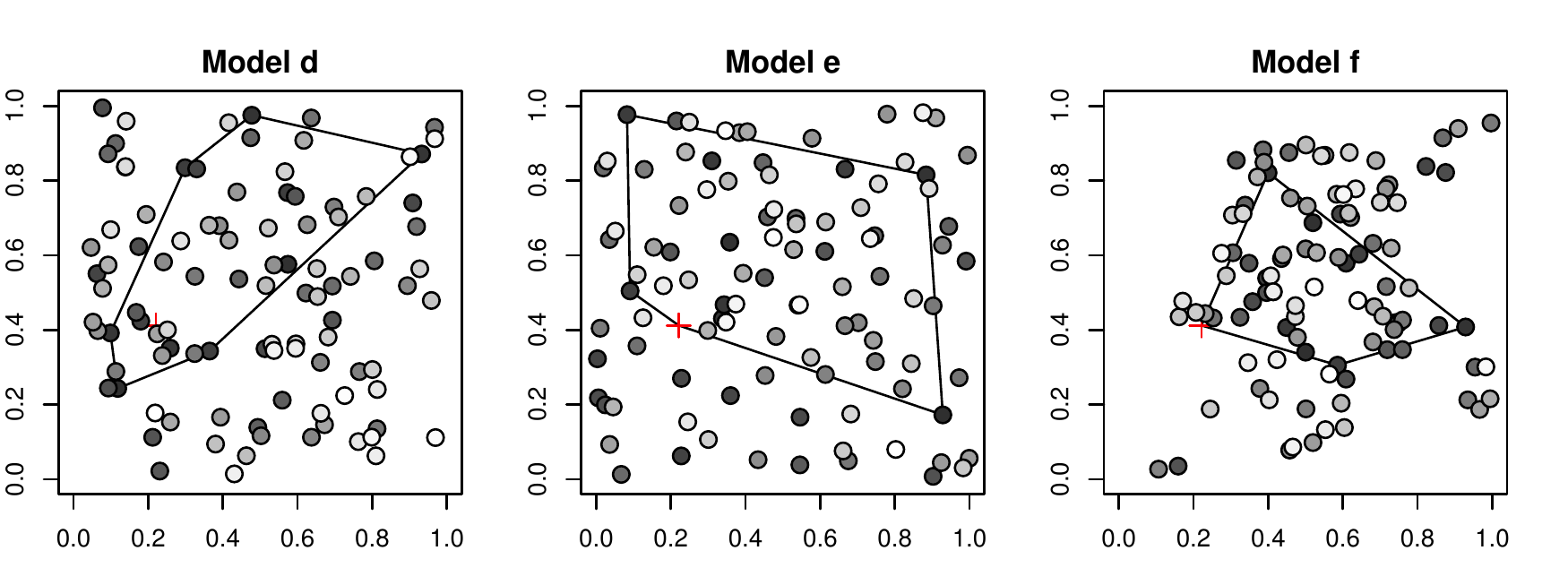}
  \caption{Simulated patterns of the three history-dependent models with recurrence self-interaction. The colour of the points denotes their order (from dark to light) and the first fixation is marked with a cross. The polygon indicates the convex hull of the first 10 points.}
  \label{fig:PointsModel1}
\end{figure}

We again simulate 19 realisations of the random walk model 
{\em d} as well as five realisations of Model {\em e} and Model {\em f}. In
Figure \ref{fig:PointsModel1} we can see that, compared with the
random walk model {\em d}, the realisation of the low recurrence model {\em e} indicates a tendency towards higher regularity, and the realisation of the high recurrence model {\em f} is clearly more clustered. Note also that Model {\em e} seems to cover the whole area quite fast compared with the other two models. 
 
\begin{figure}[!ht]
\centering
  \includegraphics[width=0.7\textwidth]{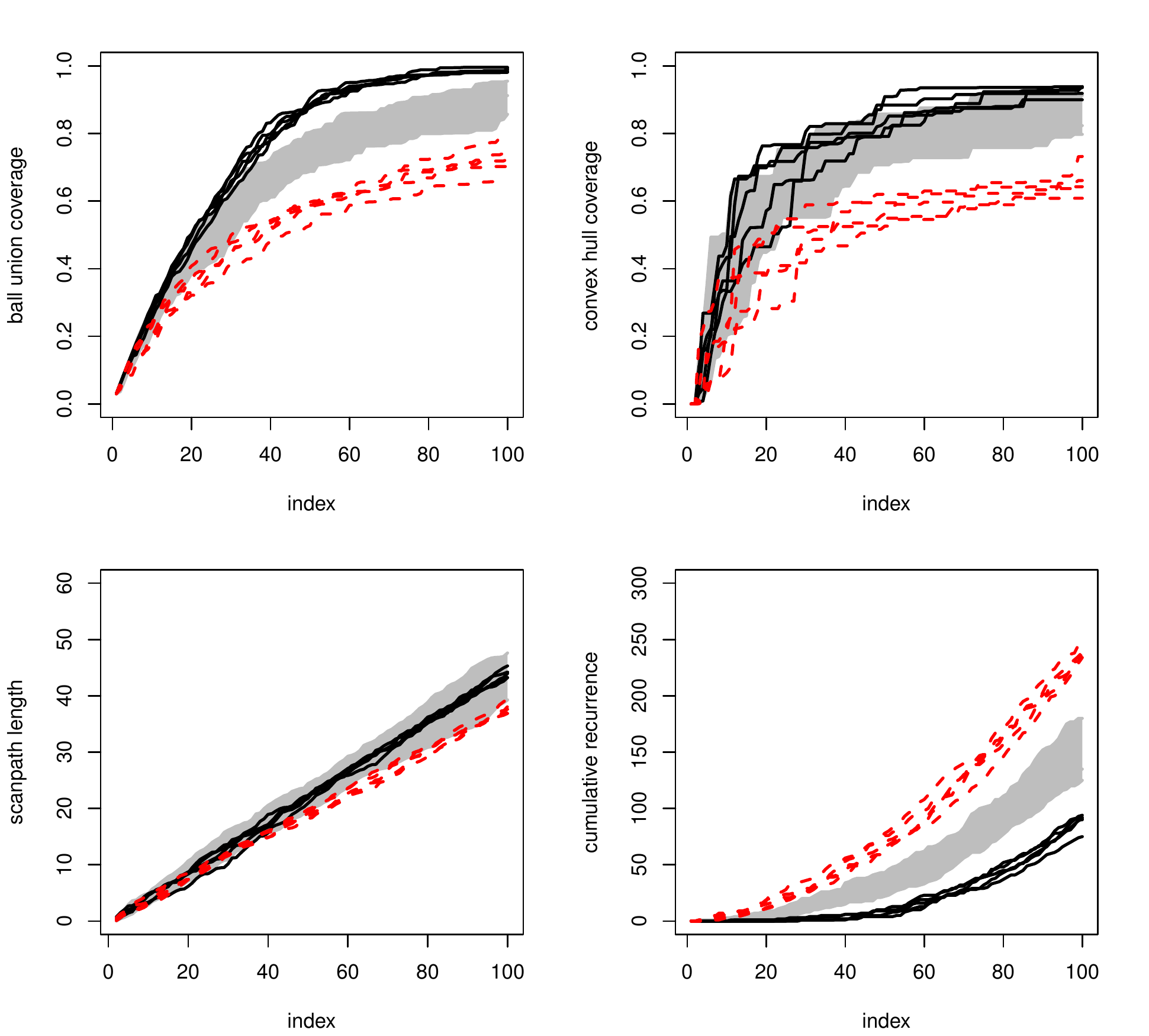}
  \caption{Ball union coverage with radius 0.1 (top left), convex hull coverage (top right), scanpath length (bottom left) and cumulative recurrence with radius 0.1 (bottom right) for the two models (five realisations of each): The low recurrence model {\em e} is marked with black solid line and the high recurrence model {\em f} is marked with red dashed line. The grey area represent the envelopes estimated from 19 realisations of the random walk model {\em d}.}
  \label{fig:SummariesModel1}
\end{figure}

The results of the functional summary statistics are depicted
in Figure \ref{fig:SummariesModel1}. The ball union coverage
describes clustering of the points, hence for the high recurrence model
{\em e} the ball union coverage curves locate
lower than for the low recurrence model {\em f}, which covers the whole window with 100 points. The convex hull
coverage curves reveal an effect similar to the ball union coverage:
the low recurrence model {\em e} almost fills the whole unit
square, whereas the high recurrence model {\em f} only fills around
60 \% of the area.

There is not much difference between the processes when
comparing the scanpath lengths for the first 40 points, but after that the
high recurrence model {\em f} makes shorter jumps on average compared with the other two models. However, the cumulative recurrence function clearly reveals that the two models differ from the random walk model: the low recurrence
model {\em e} avoids areas close to the previous included points, while the high recurrence model {\em f} favours areas near the previous included points. 

\subsection{History-adapted model with convex hull coverage \\ self-interaction}

In this example we fix the kernel width parameter $\tau = 0.3$ and pay attention to
the effect of the parameter $\kappa$ of the history-adapted model (\ref{model-spec})-(\ref{model-spec2}). The parameter $\kappa$ controls the speed of decay as a function of coverage. We again define three
history-adapted models:
\begin{itemize}
\item {\em Model g}, $\kappa = 0$ (random walk,
the kernel does not change in time). \item {\em Model h}, $\kappa =
2$ (mild clustering), which means that the process is allowed to take long
jumps at the beginning, but eventually starts to cluster.  \item {\em
Model i}, $\kappa = 4$ (fast clustering), which starts to
cluster rather quickly when the coverage increases.
\end{itemize}

The kernel function (\ref{model-spec}) here uses the convex hull
coverage, which means that in (\ref{model-spec2}) $S(\ora x_k)$ is
the area of the convex hull coverage generated by the points
$(x_1, \dots, x_k)$. Now the history-adapted model works
in such a way that at first the kernel width parameter $\tau$ is dominating and the process can make long jumps, 
but when the area of the convex hull of points approaches the size
of the window, the parameter $\kappa$ starts to affect and
produces clustering.

\begin{figure}[!ht]
\centering
  \includegraphics[width=1\textwidth]{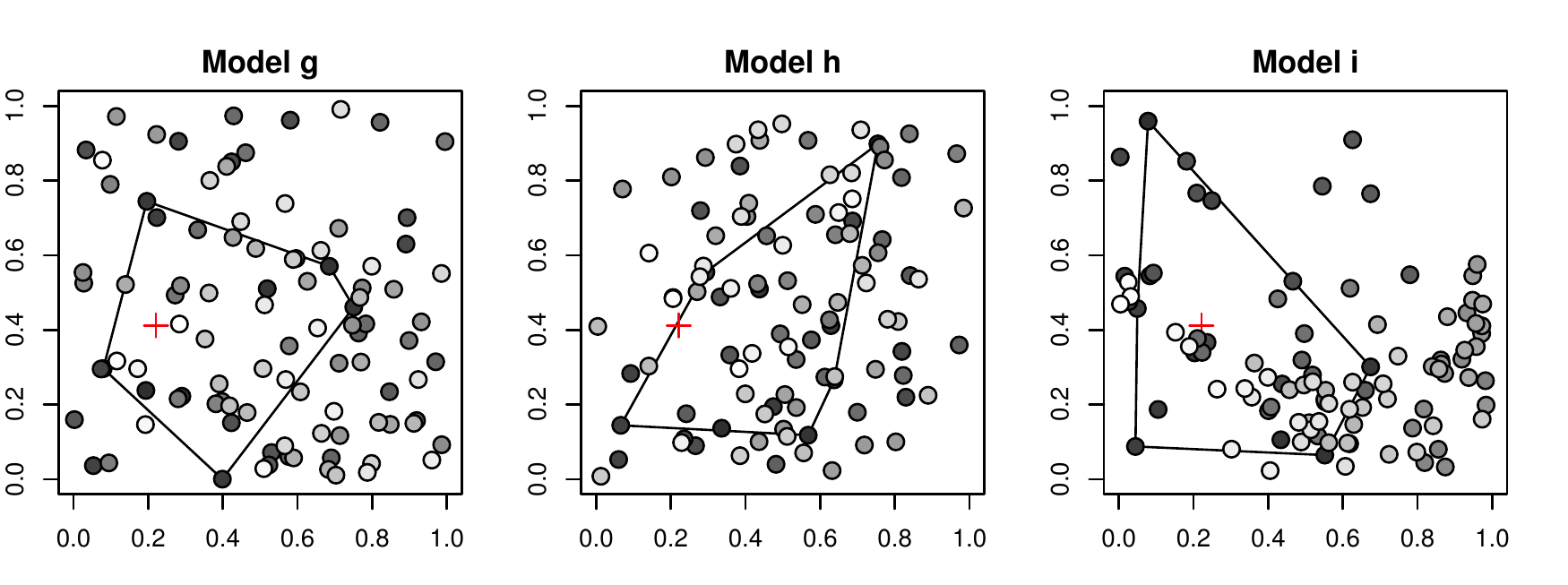}
  \caption{Simulated patterns of the three history-adapted models. The colour of the points denotes their order (from dark to light) and the first fixation is marked with a cross. The polygon indicates the convex hull of the first 10 points.}
  \label{fig:PointsModel2}
\end{figure}

\begin{figure}[!ht]
\centering
  \includegraphics[width=0.7\textwidth]{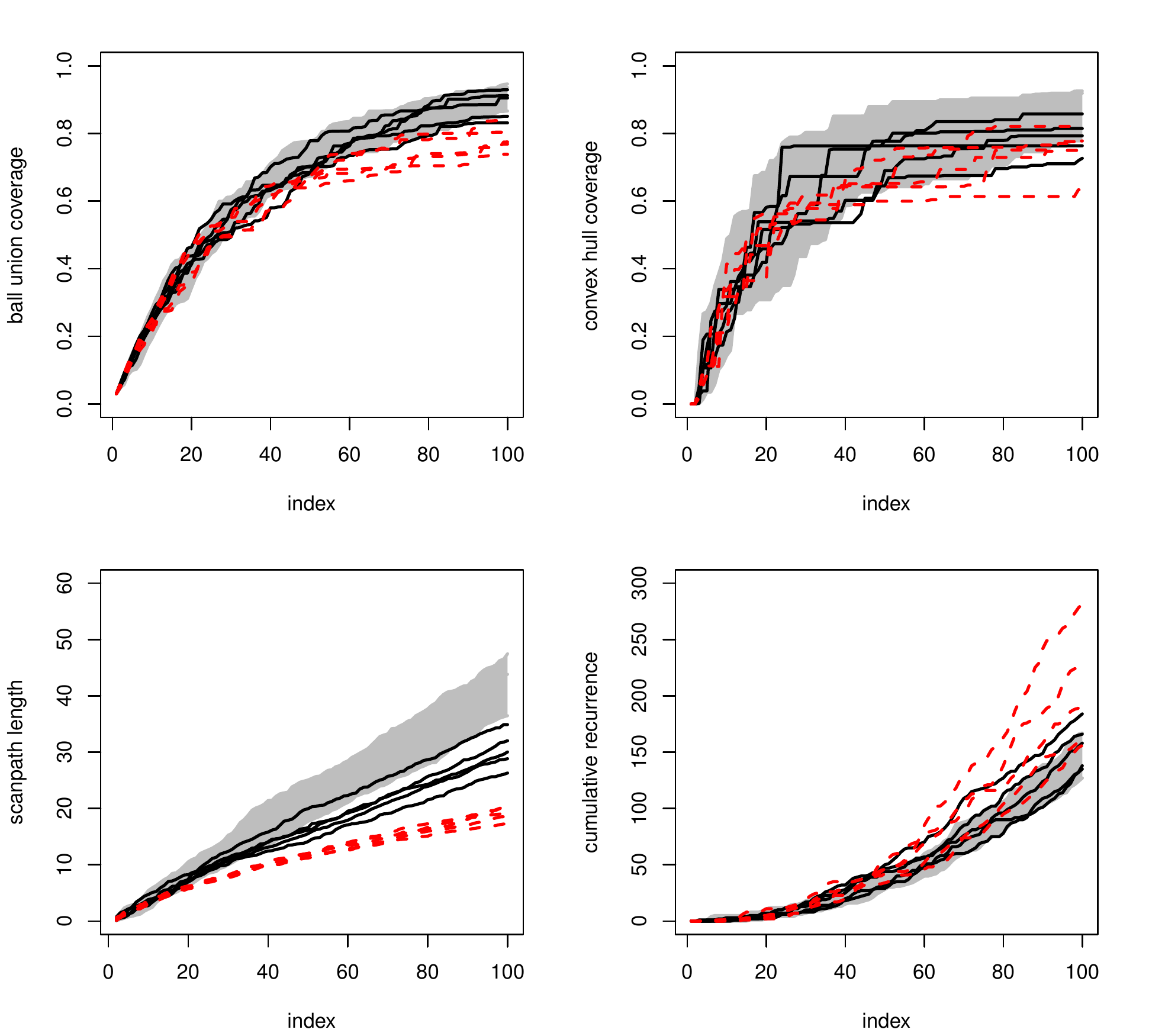}
  \caption{Ball union coverage with radius 0.1 (top left), convex hull coverage (top right), scanpath length (bottom left) and cumulative recurrence with radius 0.1 (bottom right) for the two models (five realisations of each): The mild clustering model {\em h} is marked with black solid lines and the fast clustering model {\em i} is marked with red dashed lines. The grey area represents the envelopes estimated from 19 realisations of the random walk model {\em g} used as the reference model.}
 \label{fig:SummariesModel2}
\end{figure}

In Figures \ref{fig:PointsModel2} and \ref{fig:SummariesModel2}, it can be seen that the
convex hull of the first 10 points is of the same size for
all models: the speed of coverage
seems to be similar for all processes at the beginning. The
spatial structure of the mild clustering model {\em h} and the random
walk model {\em g} are quite similar, but the points of the fast
clustering model {\em i} are clearly more clustered than the points of the other two models, and there are only a
few points in the upper half of the unit square.

The functional summary statistics are plotted in Figure
\ref{fig:SummariesModel2}. The fast clustering model {\em i} covers
the area similarly to the other two models at the beginning, but after about 50 points it starts to cluster, which can be seen as a decline of the ball union coverage summaries. 
The convex hull coverage does not reveal much difference between the models, and all the models are able to cover at least 60 \% of the window. This is due to the wide kernel ($\tau = 0.3$) which allows the processes to make long jumps at the beginning.

The summary statistic that shows the clearest
difference between the three models is the scanpath length. While
the jumps in the random walk model have time invariant transitions, the
clustering models {\em h} and {\em i} start to take shorter jumps
at some point, which is indicated by the decline in scanpath
curves. In addition, the cumulative recurrence function shows that
the fast clustering model {\em i} gathers points around the
previous ones. To conclude, the effect of the decay parameter $\kappa$ seems to fasten the clustering as a function of coverage.

\section{A case study: Black Bow by Wassily Kandinsky}

We apply the developed modelling to experimental eye movement data related
to arts in order to study self-interaction. The participants of the art experiment were inspecting six
pictures of paintings and their eye
movements were recorded. The stimulus picture was shown on the
screen with a 1024 $\times$ 768 resolution and the eye movements were
measured by the SMI iView X\texttrademark Hi-Speed eye tracker
with temporal resolution of 500 Hz. The distance between a
participant's head and the screen was about 85cm, and a forehead
rest was used in order to prevent unintentional head movements. Each
stimulus painting was shown for three minutes. The participants
were also asked to describe the moods of the painting and
their voice was recorded, but this information is not used here. 

We will focus on one painting used in the experiment, called Black Bow (1912) by Wassily Kadinsky shown in Figure \ref{fig:Points_kh5} (source of the painting:
\cite*{duchting}). We will fit four versions of the
history-dependent model with recurrence self-interaction to the
eye movement data of one subject. The goodness-of-fit of
the model is checked using the four functional summary statistics mentioned earlier, and
the best fitting model is compared with the other subjects' data in order to
conclude whether the same model fits well for all participants.

\subsection{Fitting the history-dependent rejection model with\\ recurrence self-interaction}
\label{fitting_rej_model}

We choose one subject of which eye movements are analysed and
modelled here. Because of the long inspection period (three
minutes), we decided to use only 100 first fixations of the
sequence corresponding to a 35 second time-interval, as shown in Figure \ref{fig:Points_kh5}. According to the two-stage model \citep[see e.g.][]{locher_gray_nodine} the overall impression of the scene is obtained during the first
few fixations, and then the focus turns to the presumably
interesting features. In addition, the gaze has a tendency to return to the interesting parts of the scene. Our aim is to find out
whether we can find this sort of behaviour, i.e.\ if the process
is of self-interacting type and if we can catch it with our rejection
model.

We first investigate the variation of the four functional summary
statistics related to this particular data from Kadinsky's
painting. The ball union coverage with radius of 35 pixels, convex hull
coverage, scanpath length and cumulative recurrence (radius 50) of the 20 subjects
of the experiment are presented in Figure \ref{fig:M4_results}, as a dark solid curve for the subject under study, and as grey curves for the other participants. It can be noticed
that the first 100 fixations do not cover the whole painting (the ball
union covers around 30 \% and the convex hull around 40 \% of the
target). It is typical of the eye movements that the edges of the
painting are avoided and that is why the coverage hardly ever reaches
the whole scene.

\begin{figure}[!ht]
\centering
  \includegraphics[width=0.5\textwidth]{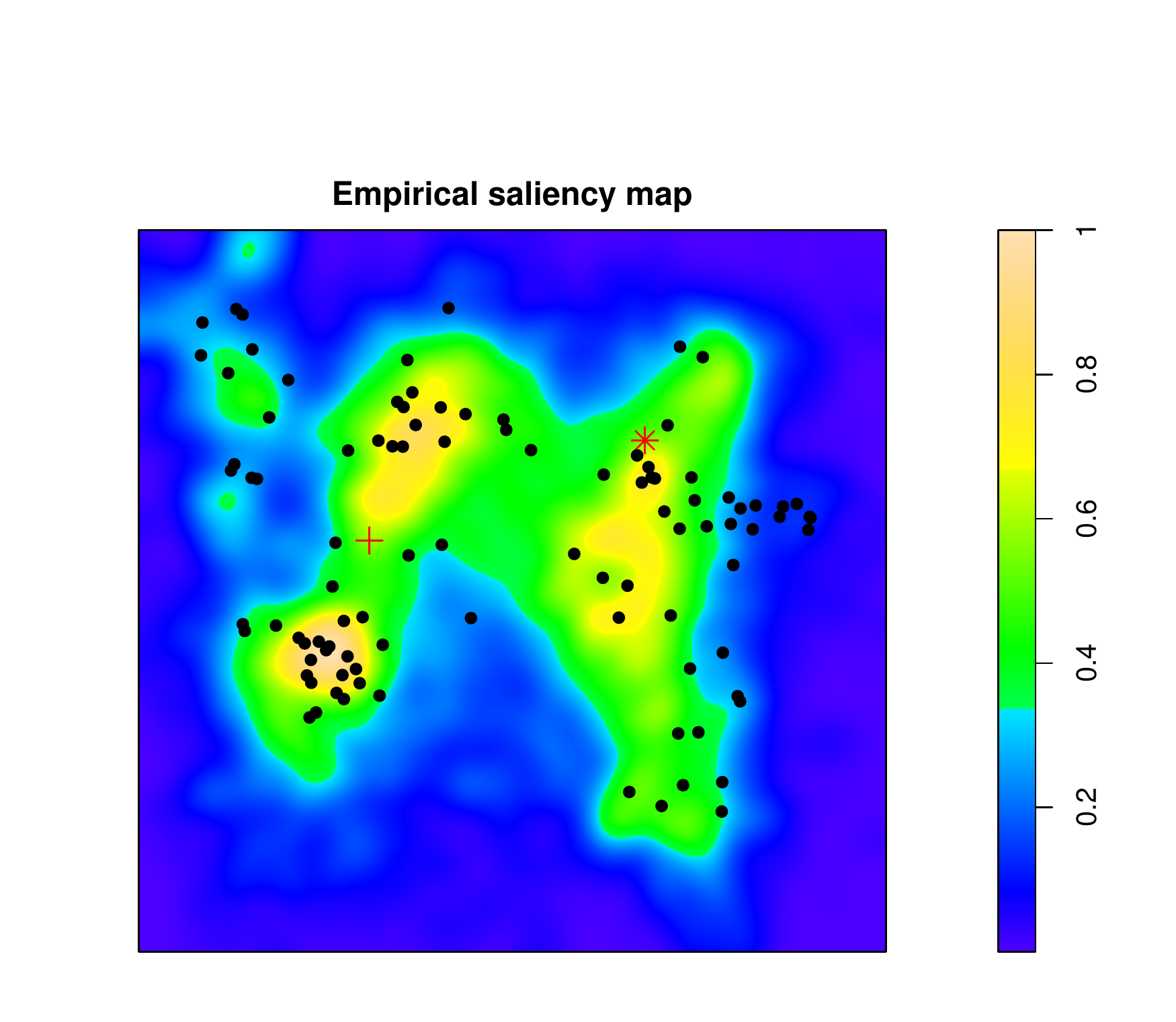}
  \caption{Empirical saliency map estimated from the fixations of all subjects excluding the one under study. The points indicate the first 100 fixation points of the particular subject under study. The first fixation is marked with a cross and the second with a star.}
  \label{fig:EmpiricalSaliencyMap_kh5}
\end{figure}

Next, we estimate the heterogeneity term
$\alpha(x)$ for the target painting. In this case, we utilize the
empirical saliency map estimated as the intensity of fixation
patterns of all the 20 subjects excluding the one under
study (a total of 9366 fixations is used for the intensity estimation). Problems associated with the estimation of $\alpha(x)$ are considered in Discussion. For technical reasons $\alpha(x)$ is scaled to have
values in $[0,1]$. The scaled saliency map together with the fixations of
the subject under study can be seen in Figure
\ref{fig:EmpiricalSaliencyMap_kh5}. This particular subject paid most attention to the areas with high
intensity, but the gaze stayed in some low intensity areas also. 

In what follows, we fit four different rejection models consisting
of three components: heterogeneity (H), contextuality (C) and
self-interaction (S). The self-interaction is assumed to be caused by
the delayed recurrence function $S_R(\ora x_{k},x),\ k = 2,\dots,n-1$ defined in (\ref{del_rec}), where the radius $r=50$ pixels is used. Therefore, we condition by the first two fixations, $x_1$ and $x_2$. All four models are submodels of (\ref{trans12}), see Table \ref{fourmodels}. 

\begin{table}[!ht]
\centering
\begin{tabular}{c|l|l}
  Model & Components & Parameter values\\
  \hline
  1 & H &  $\theta = \frac{1}{2}$, $\sigma^2$ large*\\
  2 & H, C & $\theta = \frac{1}{2}$, $\sigma^2>0$\\
  3 & H, S & $0 \le \theta \le 1$, $\sigma^2$ large*\\
  4 & H, C, S & $0 \le \theta \le 1$, $\sigma^2>0$ \\
  \hline
\end{tabular}
\caption{Components of the four rejection models. (*The value of
$\sigma^2$ should be chosen to be large enough such that the
kernel is flat in the specified window.)} \label{fourmodels}
\end{table}

Model 1 includes heterogeneity and is a binomial process in a
heterogeneous environment. When the Markovian kernel function is added (Model
2) the process is a random walk with Markovian property. Model 3
is a self-interacting process in a heterogeneous media without the contextuality effect and Model 4
contains both the Markov kernel and the self-interaction term.

The log-likelihood function for Model 4 is now
\begin{eqnarray}
\label{LL3} &\vphantom{=}&l(\sigma^2,\theta) = \sum_{k=2}^{n-1}
\log(\alpha(x_{k+1}))-\frac
1{2\sigma^2}\sum_{k=2}^{n-1}||x_k-x_{k+1}||^2 \\ &+&
\log(1-\theta)\,\sum_{k=2}^{n-1} \I_{\{S_R(\ora x_{k},x_{k+1})
=0\}}(x_{k+1}) + \log(\theta)\,\sum_{k=2}^{n-1} \I_{\{S_R(\ora
x_{k},x_{k+1})
\ge 1\}}(x_{k+1}) \nonumber\\
&-& \sum_{k=2}^{n-1}\log\int\displaylimits_W\alpha(u)\,e^{-\frac
1{2\sigma^2}\,||x_k-u||^2}\, ((1-\theta)\I_{\{S_R(\ora
x_{k},u)=0\}}+\theta\,\I_{\{S_R(\ora
x_{k},u)\ge
1\}}) \,{\rm d}u. \nonumber
\end{eqnarray}
The likelihood for Model 1 is just the first term on the right
hand side of equation (\ref{LL3}). For Model 2, the likelihood is
obtained choosing $\theta=\frac{1}{2}$ in (\ref{LL3}) and for
Model 3 choosing  $\sigma^2$ to be large (i.e.\ large enough such that the kernel is flat in the specified window). The parameters are estimated using the profile-likelihood method with few iterative steps, see Section 2.3.1. for details. We used numerical integration for computing the normalization term of the log-likelihood (\ref{LL3}), last row, and a grid of $(60,80, \dots, 400)$ and $(0.05, 0.10, \dots, 0.95)$ for maximizing the likelihood.

Model 1 includes only the empirical saliency map and we do not
have to estimate any parameters. For Model 2 we get ${\hat \sigma}
= 180$, and for Model 3 ${\hat \theta} = 0.75$. The parameter
estimates for Model 4 are  ${\hat \sigma} = 180, {\hat \theta} =
0.70$. Note that for a random walk model we should have $\theta =
0.50$; hence there are recurrence features involved in this data.

\subsubsection{Model comparisons} 

\begin{figure}[!ht]
\centering
  \includegraphics[width=0.75\textwidth]{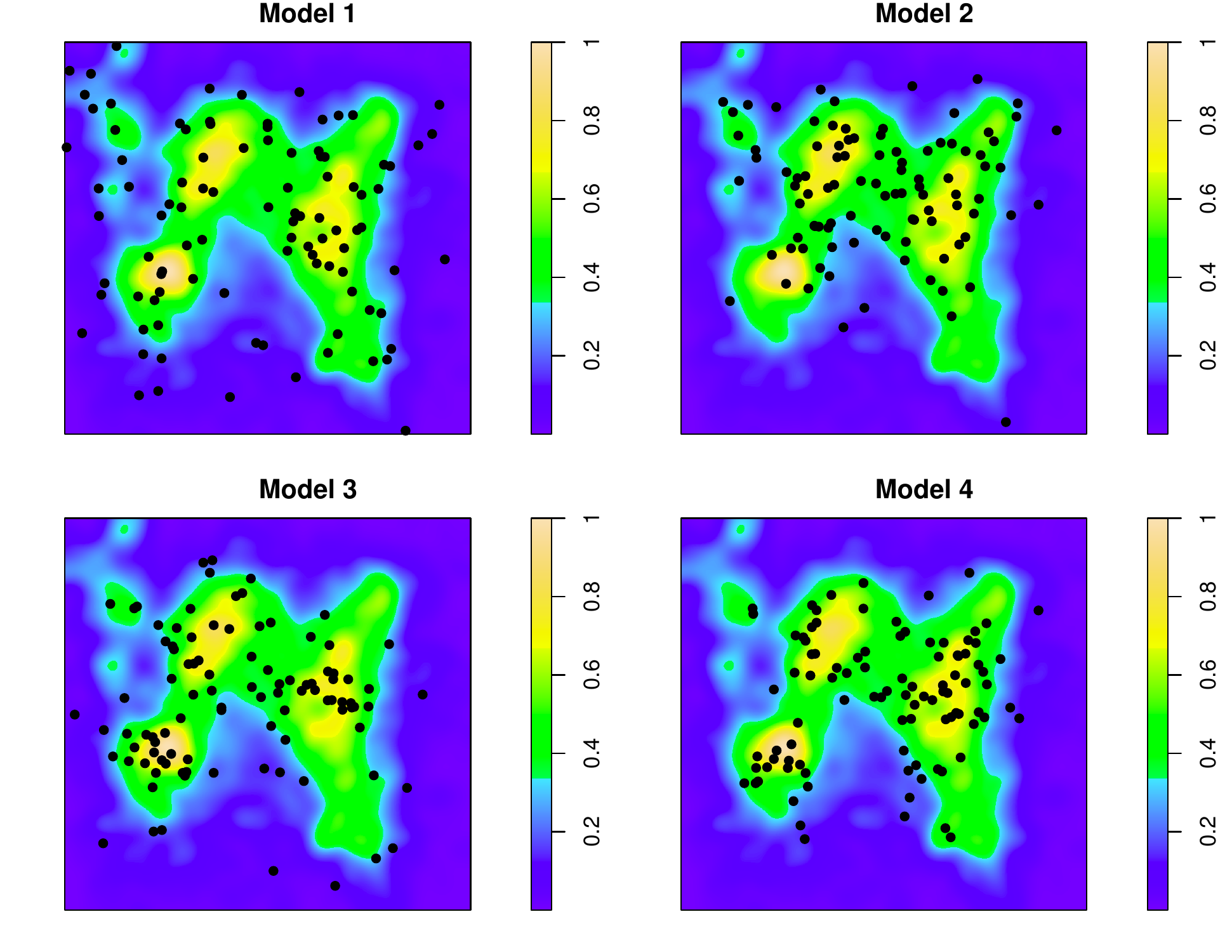} 
  \caption{Simulated realisations of the four models fitted to the eye movement data of the particular subject under study, overlaid with the empirical saliency map based on data from the other subjects. The first two points are fixed and marked with a cross and a star, respectively. } 
  \label{fig:Fitted_models}
\end{figure}

\begin{figure}[!ht]
\centering
  \includegraphics[width=0.75\textwidth]{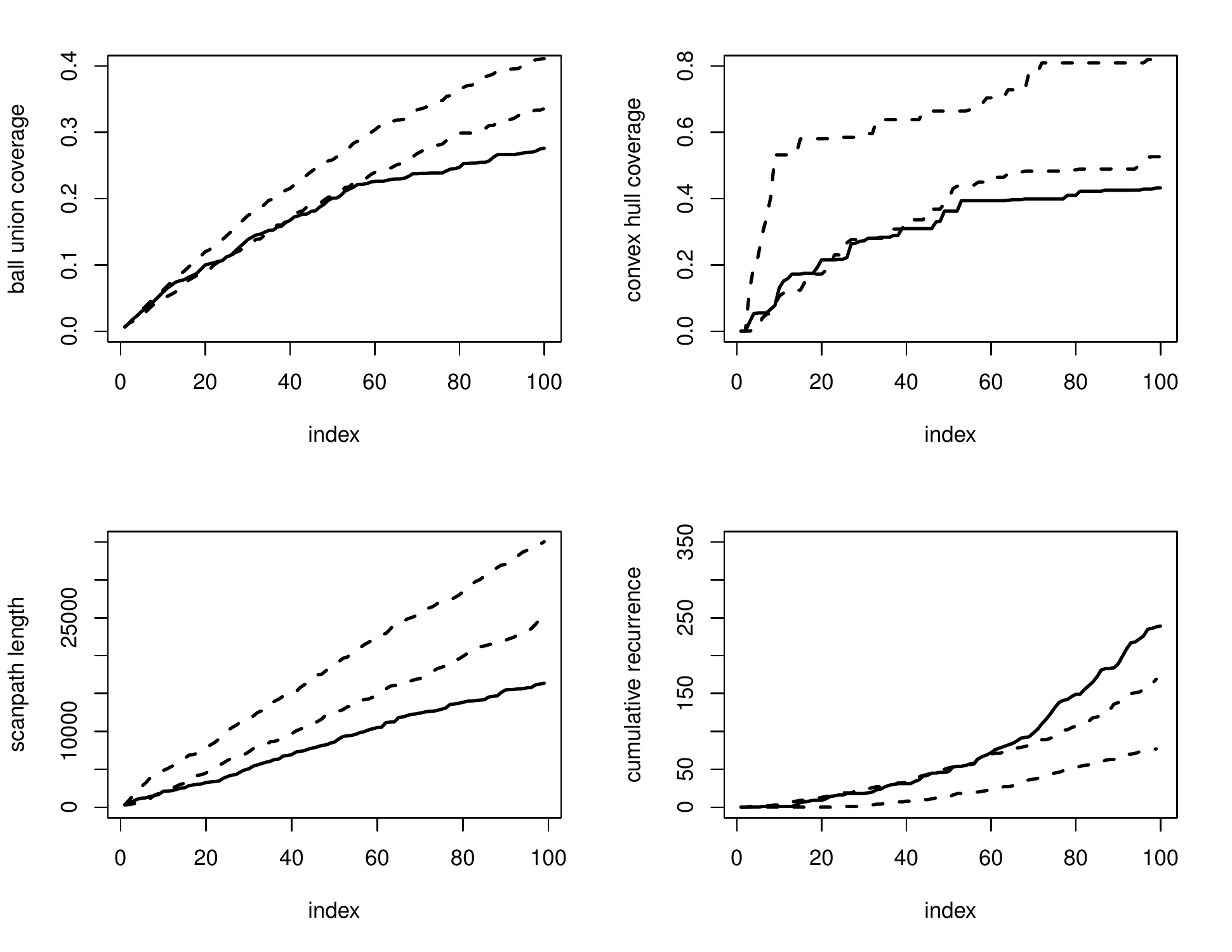}
  \caption{Ball union coverage with radius 35 (top left), convex hull coverage (top right), scanpath length (bottom left) and cumulative recurrence with radius 50 (bottom right) for the subject under study (solid line). Dashed lines represent pointwise envelopes estimated from 99 simulations of Model 1.}
  \label{fig:M1_results}
\end{figure}

\begin{figure}[!ht]
\centering
  \includegraphics[width=0.75\textwidth]{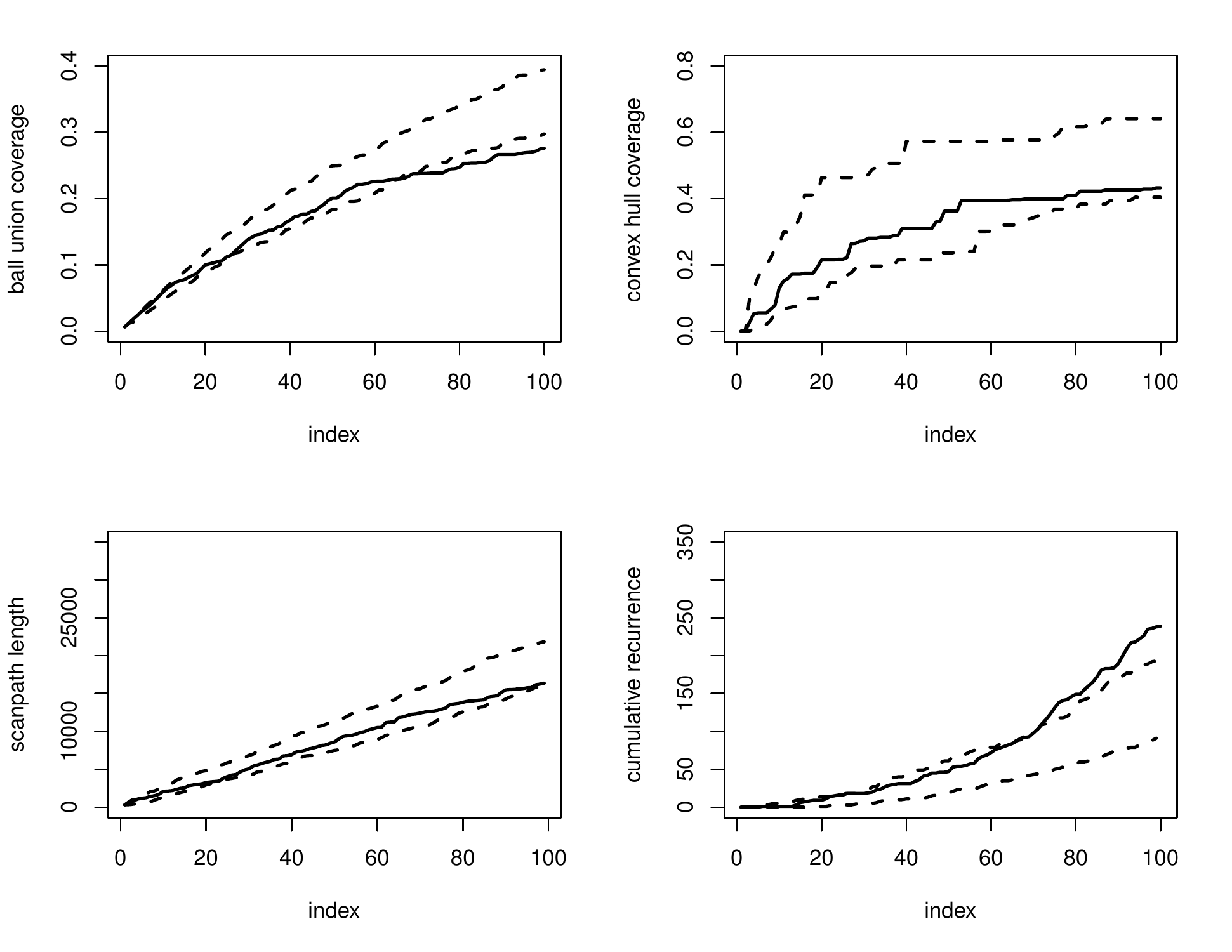} 
  \caption{Ball union coverage with radius 35 (top left), convex hull coverage (top right), scanpath length (bottom left) and cumulative recurrence with radius 50 (bottom right) for the subject under study (solid line). Dashed lines represent pointwise envelopes estimated from 99 simulations of Model 2.}
  \label{fig:M2_results}
\end{figure}

\begin{figure}[!ht]
\centering
  \includegraphics[width=0.75\textwidth]{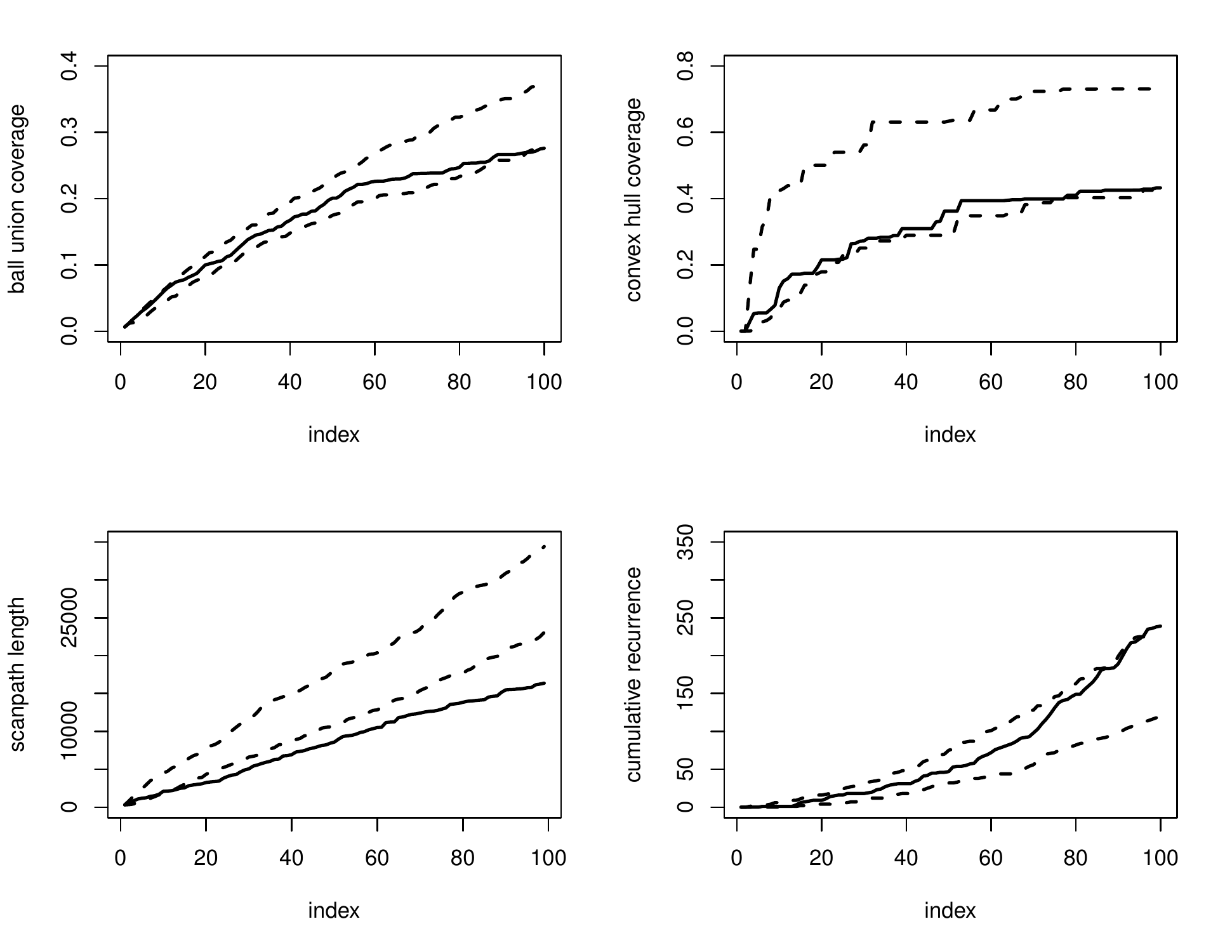} 
  \caption{Ball union coverage with radius 35 (top left), convex hull coverage (top right), scanpath length (bottom left) and cumulative recurrence with radius 50 (bottom right) for the subject under study (solid line). Dashed lines represent pointwise envelopes estimated from 99 simulations of Model 3.}
  \label{fig:M3_results}
\end{figure}

\begin{figure}[!ht]
\centering
  \includegraphics[width=0.75\textwidth]{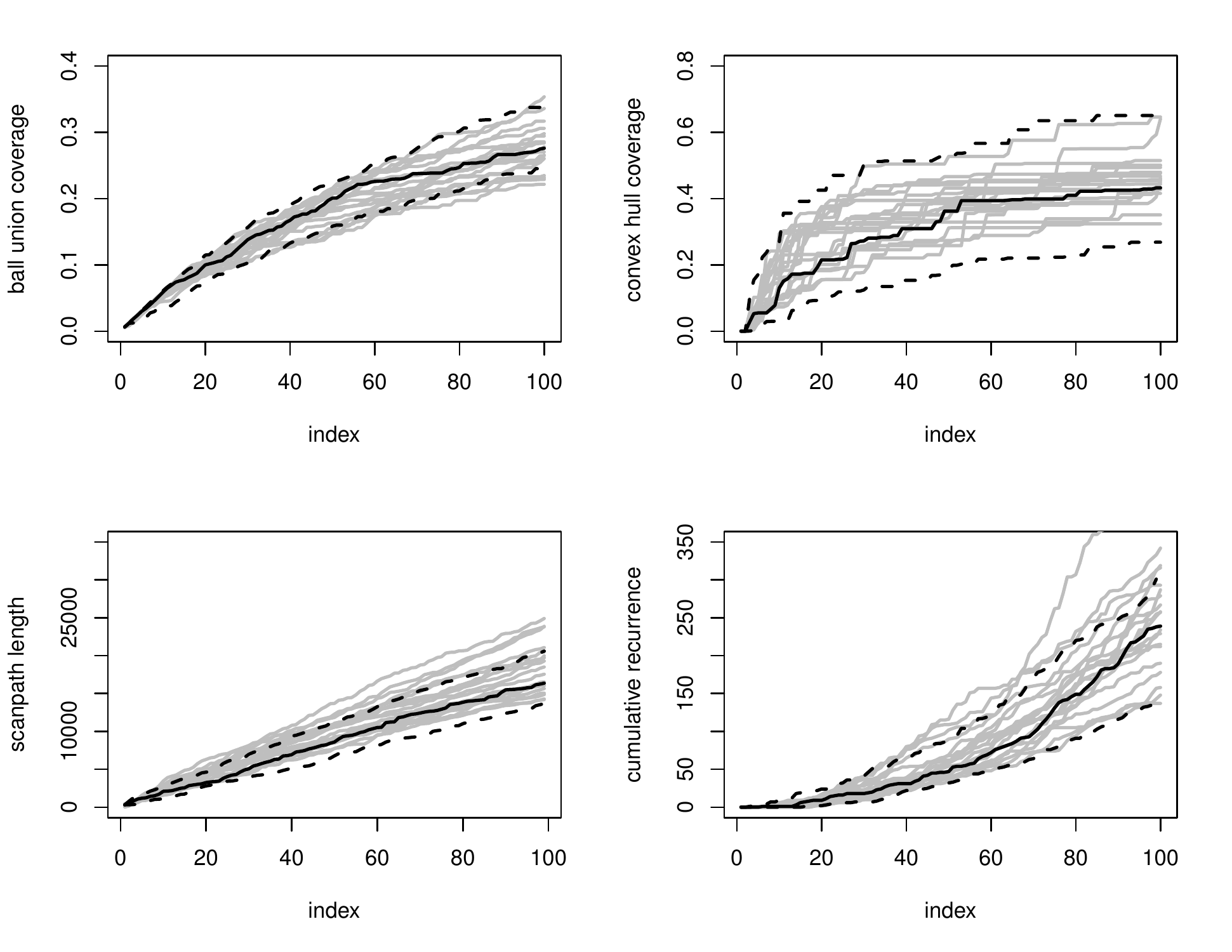} 
  \caption{Ball union coverage with radius 35 (top left), convex hull coverage (top right), scanpath length (bottom left) and cumulative recurrence with radius 50 (bottom right) for the subject under study (black solid line) and for all other subjects (grey solid lines). Dashed lines represent pointwise envelopes estimated from 99 simulations of Model 4.}
  \label{fig:M4_results}
\end{figure}

Assessing the goodness-of-fit of the models is here done by estimating the four summary
statistics mentioned earlier from the data and from 99 simulated realisations of
the fitted models. When simulating the model, we condition on the observed values of the first two
fixations in order
reduce variation right at the beginning of the process. One
simulated realisation of each model can be seen in Figure
\ref{fig:Fitted_models}, and the summary statistics estimated from the data with
pointwise envelopes estimated from the simulations in Figures
\ref{fig:M1_results} -- \ref{fig:M4_results}. Note that the ball union and convex hull coverages are here presented with respect to the size of the window, hence they obtained values in $[0,1]$. 

For Model 1 all summary statistics show poor fit (Figure
\ref{fig:M1_results}). Compared with the data this model covers the target area too fast, takes too long jumps according to the scanpath
length, and goes to areas with too few points according to the
cumulative recurrence function. As a conclusion, the heterogeneity component alone does not describe the data set well enough even though the spatial heterogeneity is followed rather well (see Figure
\ref{fig:Fitted_models} upper left).

Model 2 seems to perform slightly better than Model 1. The summaries estimated
from the data set stay inside the simulated pointwise envelopes, except the ball
union coverage and cumulative recurrence function after the first 70
points (Figure \ref{fig:M2_results}). These findings indicate that data begin to cluster at the end of the inspection period, but the model does not carry that effect. 

Model 3 includes heterogeneity and self-interaction, but not
contextuality, which is related to the length of the jumps the process
makes. As a result, this model seems to jump too much compared with data, since the estimated scanpath length summary is at odds with
the simulated pointwise envelopes (Figure \ref{fig:M3_results}). The marginal spatial structure
looks slightly more clustered than Model 1 and Model 2 predict (Figure \ref{fig:Fitted_models}).

Model 4 includes all the three effects and seems to be in good
agreement with data: all four summary statistics estimated for
the subject under study stay within the simulated pointwise envelopes, see
Figure \ref{fig:M4_results}. It seems that this model is able to
catch the nature of this eye movement process fairly well. The estimated
parameter value ${\hat \theta} = 0.75$ indicates that the locations nearby
the previous points (excluding the most recent point) are favoured, which is a cause of spatial clustering. We conclude that the random walk model does not seem to be
a good model for these data, but there is self-interaction due to the recurrence
involved: the eye movement process seems to favour areas close to
previous fixations.

\subsubsection{Population level comparison}

We have been able to describe the variation in the eye movement sequence of an individual by using the rejection model with recurrence-based weighting. In order to investigate the generality of the suggested models, we make comparisons at the population level using all 20 subjects. As can be
seen in Figure \ref{fig:M4_results}, the envelopes of Model 4 seem
to cover the convex hull and ball union coverage
curves of the subjects rather well. The scanpath length does not cover the curves as well: there are subjects
whose gaze makes longer jumps at the beginning of the eye movement process
than the fitted model predicts. Furthermore, the envelopes of the
cumulative recurrence function cover almost all the curves, but there is one exceptional subject, whose
fixations are strongly clustered after the 70th fixation.

In order to further describe the variation in the data set related to
self-interaction, we fitted Model 4 for the first 100
fixations of each subject separately. The estimates of the parameters
$\sigma$ and $\theta$ can be seen in Table
\ref{table_of_parameter_estimates}. The 90 \% bootstrap confidence intervals for the parameter estimates are calculated from 20 realisations of the fitted model. When the parameter $\sigma$ is
large, the model allows jumps over the observation window, and then the
self-interaction parameter $\theta$ dominates. Large $\theta$ indicates strong spatial clustering. When $\sigma$ is small, one needs to take multiple jumps in order to cross the whole target window.

\begin{table}[!ht]
\centering
\begin{tabular}{|c|c|rc|c|rc|}
  \hline
  Subject id & $\hat{\sigma}$ & \multicolumn{2}{c|}{90 \% CI for $\hat{\sigma}$} & $\hat{\theta}$  & \multicolumn{2}{c|}{90 \% CI for $\hat{\theta}$}\\
  \hline
    1 & 240 & 200 & 280  & 0.75 & 0.70 & 0.85 \\
    2 & 180 & 160 & 200 & 0.55 & 0.45 & 0.65 \\
    3 & 160 & 140 & 180 & 0.65 & 0.60 & 0.75 \\
    4 & 160 & 140 & 180 & 0.80 & 0.75 & 0.85 \\
    {\bf 5} & {\bf 180} & {\bf 140} & {\bf 200} & {\bf 0.70} & {\bf 0.65} & {\bf 0.85} \\
    6 & 180 & 140 & 200 & 0.85 & 0.80 & 0.95 \\
    7 & 160 & 140 & 180 & 0.70 & 0.65 & 0.80\\
    8 & 240 & 200 & 280 & 0.70 & 0.65 & 0.80  \\
    9 & 160 & 140 & 180 & 0.65 & 0.60 & 0.75 \\
    10 & 160 & 140 & 180 & 0.65 & 0.60 & 0.75 \\
    11 & 200 & 180 & 220 & 0.70 & 0.65 & 0.80\\
    12 & 260 & 240 & 320 & 0.85 & 0.80 & 0.90 \\
    13 & 280 & 220 & 340 & 0.70 & 0.60 & 0.85 \\ 
    14 & 220 & 200  & 240 & 0.70 & 0.60 & 0.75 \\
    15 & 220 & 180 & 260 & 0.55 & 0.45 & 0.60 \\
    16 & 200 & 160 & 220 & 0.70 & 0.65 & 0.80 \\
    17 & 160 & 140 & 180 & 0.70 & 0.60 & 0.80 \\
    18 & 340 & 220 & 420 & 0.80 & 0.75 & 0.90 \\
    19 & 140 & 120 & 160 & 0.70 & 0.65 & 0.80 \\
    20 & 280 & 200 & 340 & 0.80 & 0.70 & 0.85 \\
  \hline
\end{tabular}
\caption{Estimated parameters $\hat{\sigma}$ and $\hat{\theta}$ of
Model 4 with their confidence intervals for all subjects. The subject under closer study is number 5 (bold).} \label{table_of_parameter_estimates}
\end{table}

In Figure \ref{fig:M4_results}, we have plotted the estimated summary statistics for all subjects together with the pointwise envelopes based on the model fitted to subject 5. For subjects 13, 18 and 20, the scanpath length curves clearly exceed the envelopes of the Model 4 fitted for subject 5. For each of them, the fitted value of the parameter $\sigma$ in Model 4 is over 280. This indicates that the process is allowed to make long jumps resulting in longer scanpaths. However, for these subjects the recurrence parameter $\theta$ varies from 0.70 to 0.80 and does not differ much from the recurrence parameter of subject 5.

For subject 6, the cumulative recurrence curve is way above the envelopes, and the estimated recurrence parameter is 0.85 indicating strong clustering. The other subjects, whose cumulative recurrence curve is outside the envelopes, are 4, 12, 18 and 20, and for all of them the recurrence parameter is over 0.80. As a conclusion, there is some variation related to the clustering effect of the points between these subjects. However, for each subject the estimated recurrence parameter differs from 0.5 meaning that the random walk is not a suitable model. 

\section{Discussion}
In this paper, we develop advanced data-analytical tools for extracting
information from eye movement sequences, which are needed in various areas of application utilizing eye tracking \citep*[see e.g.][]{rayner_2009}. Our objective is to create simple but flexible dynamic stochastic
models by employing mechanisms which use the whole history of
the sequence in each gaze jump in order to capture features of
learning during the experiment.

Heterogeneity of the scene, contextuality of subsequent fixations, and self-interaction of eye movements are elements that affect the eye movement process.
We present a sequential spatial point process
approach which includes these three effects, the self-interaction
being new in this context and is interpreted as a learning
effect. This leads to what in probability theory is called
self-interacting processes, which are generalizations of random walks in heterogeneous
media. Although self-interacting random walks are well established in
mathematics, physics and animal ecology, our reasoning here is slightly different. We study how the process evolves at an early stage of an eye movement sequence whilst, e.g.\ in
mathematics, the long term behaviour is of interest.
Such processes are analytically difficult,
even intractable, but their simulation is basically straightforward.

After having constructed a new model, we need model fitting (estimation), evaluation of
goodness-of-fit (model criticism) and simulation algorithms
for various inferential purposes. Here, we suggest a
likelihood approach for the new processes which is used in parameter
estimation. It is not possible to obtain analytical results or
use asymptotical reasoning.  Instead, we compute the likelihood using simulation which allows us to make exact inference in the sense that it does not depend on the size of data and which includes a boundary effect correction. In doing this, we enlarge the applicability of spatial statistics and the
likelihood inference to a new area of applications.
The processes can be used to make inference on the structure of data, including self-interaction, and to deduce uncertainties
in conventional and new functional data summaries such as
scanpath length and recurrence function.

In this paper, we are interested in the dynamics of the eye movement process. The main question is whether there is self-interaction present in a given eye movement sequence and whether we can detect it using our new modelling. We focus on the beginning of an eye movement process, since, according to the two-stage model by Locher and colleagues \citep[see e.g.][]{locher}, the gist of the scene is established during the early fixations. Our history-dependent rejection model is, in fact, able to observe self-interaction in these particular data, although there is large within-subject variation.

Our models utilize stochastic geometry in creating self-interaction 
caused either by coverage (how much of the scene is covered and how fast) 
or by recurrence (how much the process favours points nearby the previous points),
both of which have justifications in eye movement literature. 
Functional summary statistics are needed for checking the goodness-of-fit of
a fitted model, as well as for describing the structural components 
of the sequence. We use four summary statistics: convex hull coverage,
ball union coverage, scanpath length, and cumulative recurrence. 
Several summary statistics are needed since none of these four was 
able to alone distinguish between the models in our simulation study. 
We found that the rejection model with convex hull coverage can be separated 
from the random walk by the scanpath length, whereas the rejection model 
with recurrence can be distinguished from random walk with the coverage 
measures. The scanpath length and cumulative recurrence are needed with 
the history-adapted model for defining the speed of spatial clustering.

We have illustrated two tractable process constructions for
self-interaction, namely, history-based independent thinning
and history-dependent transitions. These constructions are very different, and their use depends on the problem
and the data set. The two developed models are rather simple but can easily be extended. Other constructions are also possible, such as the heterogeneous mixture model, where the Markov kernel
$K(\ora x_k,x)$ is replaced by
\[
p(\ora x_k)\,K_1(x_k,x)+(1-p(\ora x_k))\,K_2(x_k,x)\,.
\]
Here, $K_1$ and $K_2$ are two kernel functions where the
choice $K_2(x_k,x)$ could be uniform in $W$ or proportional to $\alpha(x)$,
for example, and the mixture factor $p(\ora x_k)$ depends on the
history of the sequence. Although simulation of such a model
is straightforward, the associated inference is computationally demanding.

We have restricted our approach to sequential spatial point processes,
mainly due to their tractability. However, this approach is a
bridge to spatio-temporal models that would take fixation durations into account.
For a separable spatio-temporal model, the spatial effect and time
dynamics are multiplicative in the likelihood. A sequential spatial point process
model can be used as a building block: if an order-dependent spatial model
is available, the inclusion of time dynamics is straightforward,
because inference on the ordered spatial aspect and fixation
durations can be performed independently. If a preferred summary statistic contains information on the fixation durations, then a spatio-temporal model should be used instead of the sequential point process. This extension is a subject for a separate study.

The estimation of the heterogeneity component $\alpha(x)$ is an issue not fully considered here. In the second order analysis of point patterns the 
first and second order components are not estimable from one observed 
point pattern without further information. \citet{diggle2007second} suggest 
two alternatives, which are the use of a parametric model for the intensity 
(heterogeneity) or, alternatively, the utilization of replications
for the intensity estimation. In the sequential context the situation
is similar. In our experiment sequences measured from several 
participants are available and are independent of the particular sequence
under study. When using this information in the estimation of the
heterogeneity component, the problem is that also these auxiliary control 
sequences are serially correlated leading to extra clustering at the sequence
level. We assume that this effect is not as serious as in the
intensity estimation from the case data only. When using auxiliary sequences, we have assumed that these sequences contain
information which origins mainly from the target common to all participants
and to the case under study and measures the wanted heterogeneity.

An improvement would be to sample fixations from each of the 
auxiliary fixation sequences instead of using all the fixations as we did here. Merging these sampled fixations gives a 
point pattern which is then used in the estimation of $\alpha(x)$ 
using the kernel method. This will further reduce the effect of serial 
correlation. An alternative improvement is based on the case sequence under study by using the fitted model (containing
both contextuality and self-interaction and a prefixed $\alpha(x)$) 
to compute the inverse of the transition probability for each 
fixation. These weights can then be used in the estimation of 
$\alpha(x)$ by the weighted kernel method. The procedure can be 
iterated. The estimation of $\alpha(x)$ is discussed in \citet{barthelme_etal, engbert_2015}. The reliable estimation of $\alpha(x)$ with open questions is a topic 
for a subsequent study.

Another issue concerns the parameter estimation.
Here, we suggest the profile likelihood or discretized coordinate descend algorithm
for maximum likelihood using forward simulation. This early
experimenting shows that by this method it is possible to separate the effect
of self-interaction and present confidence intervals for parameter 
estimates and confidence envelopes for chosen summary
statistics. We know that approximative inference, being computationally much faster, is also a
possibility and would be very important in a methodological
toolbox. The extensive experimenting with approximative 
inference related to this is a future task.

\section*{Acknowledgements}

The authors would like to thank the three anonymous reviewers for their valuable comments, which helped in improving the manuscript. The authors are also grateful to Aila S\"arkk\"a for the useful
comments and suggestions, to Tuomas Rajala for his help with the computational issues, and to Professor Pertti Saariluoma, Sari Kuuva, Mar\'ia
\'Alvarez Gil, Jarkko Hautala, and Tuomo
Kujala for providing the data set. 

The second author has been financially supported by the Finnish
Doctoral Programme in Stochastic and Statistics and by the Academy of Finland (Project number 275929). 

\bibliography{refs2}

\end{document}